\documentclass[%
 reprint,
superscriptaddress,
 amsmath,amssymb,
 aps,
 prx,
]{revtex4-2}
\usepackage{amsmath}
\usepackage{graphicx}% Include figure files
\usepackage{dcolumn}% Align table columns on decimal point
\usepackage{bm}% bold math
\usepackage{color}
\usepackage{slashed}
\usepackage[colorlinks,allcolors=cyan]{hyperref}
\usepackage{comment}
\usepackage[normalem]{ulem} % Required for striking out text
\usepackage{color}           % Required for the red/blue coloring

\usepackage{physics}
\usepackage[caption=false]{subfig}
\usepackage{ragged2e}
\DeclareCaptionOption{justified}[]{\caption@setjustification{justified}}
\usepackage[capitalise]{cleveref}

\newcommand{\Zt}{\mathbb{Z}_2}
\newcommand{\rr}{{\vb{r}}}

\newcommand{\beq}{\begin{equation}}
\newcommand{\eeq}{\end{equation}}

\begin{document}

\title{``Odd" Toric Code in a tilted field: Higgs-confinement multicriticality, spontaneous self-duality symmetry breaking, and valence bond solids}

\author{Umberto Borla}
\affiliation{Racah Institute of Physics, The Hebrew University of Jerusalem, Givat Ram, Jerusalem 91904, Israel}
\affiliation{Max Planck Institute of Quantum Optics, 85748 Garching, Germany}
\affiliation{Munich Center for Quantum Science and Technology (MCQST), 80799 Munich, Germany}
\author{Ayush De}
\affiliation{Racah Institute of Physics, The Hebrew University of Jerusalem, Givat Ram, Jerusalem 91904, Israel}
\author{Snir Gazit}
\affiliation{Racah Institute of Physics, The Hebrew University of Jerusalem, Givat Ram, Jerusalem 91904, Israel}
\affiliation{The Fritz Haber Research Center for Molecular Dynamics, The Hebrew University of Jerusalem, Jerusalem 91904, Israel}
\date{\today}

\begin{abstract}
We investigate the quantum phase diagram of an ``odd'' variant of the two-dimensional Ising Fradkin--Shenker model, characterized by a uniform background of static $e$ and $m$ charges. Using large-scale tensor network and exact diagonalization methods, we determine the topology of the phase diagram, identifying an ``odd'' deconfined phase, confinement- and Higgs-induced valence bond solids (VBS), and a trivial paramagnet. Most notably, we uncover an exotic multicritical point along the self-dual line, where electric and magnetic excitations are related by an enriched $\mathbb{Z}_2$ duality. This transition is marked by the simultaneous onset of confinement, Higgs condensation, translational symmetry breaking, and spontaneous duality symmetry breaking. Within our numerical accuracy, the transition appears continuous, involving the softening of excitation gaps for $e$ and $m$ anyons at finite momentum. At intermediate couplings, we further identify VBS phases with enlarged unit cells, potentially indicating frustration-induced crystalline order beyond commensurate limits.

\end{abstract}

\maketitle

\section{Introduction}

Quantum critical phenomena, which cannot be understood through Landau's theory of symmetry breaking, are a cornerstone of modern condensed matter physics \cite{wenbook,Fradkin2013, sachdev_2023}. A particularly fertile playground for the study of such unconventional phases of matter is given by low-dimensional lattice gauge theory (LGT) coupled to matter fields \cite{Kogut_1979}. The presence of local gauge symmetries provides access to phases and phase transitions that cannot be captured via local symmetry-based observables \cite{Elitzur_1975}. More concretely, these include topologically ordered phases \cite{wilczek1990fractional,wenbook}, confinement and Higgs transitions \cite{Fradkin_1979}, some of which are interleaved with conventional symmetry breaking \cite{mila2012,gazit2017,gazit2018,Scaffidi_2021}. The gauge theory language has physical applications in the context of condensed matter lattice systems, where it serves as an emergent low-energy description of fractionalized phases of matter and spin liquids \cite{wilczek1990fractional,savary2016quantum,sachdev_2023}. In parallel, LGT and related spin models in one and two spatial dimensions have garnered considerable attention as initial steps toward digital and analog quantum simulations of the Standard Model \cite{barbiero2019, Homeier_2021, Lumia_2022, Zohar_2022, Homeier2023, Reinis2023}.

\begin{figure}[h!]
    \captionsetup[subfigure]{labelformat=empty}
    \subfloat[\label{subfig:pd}]{}
	\includegraphics[width=0.95\linewidth]{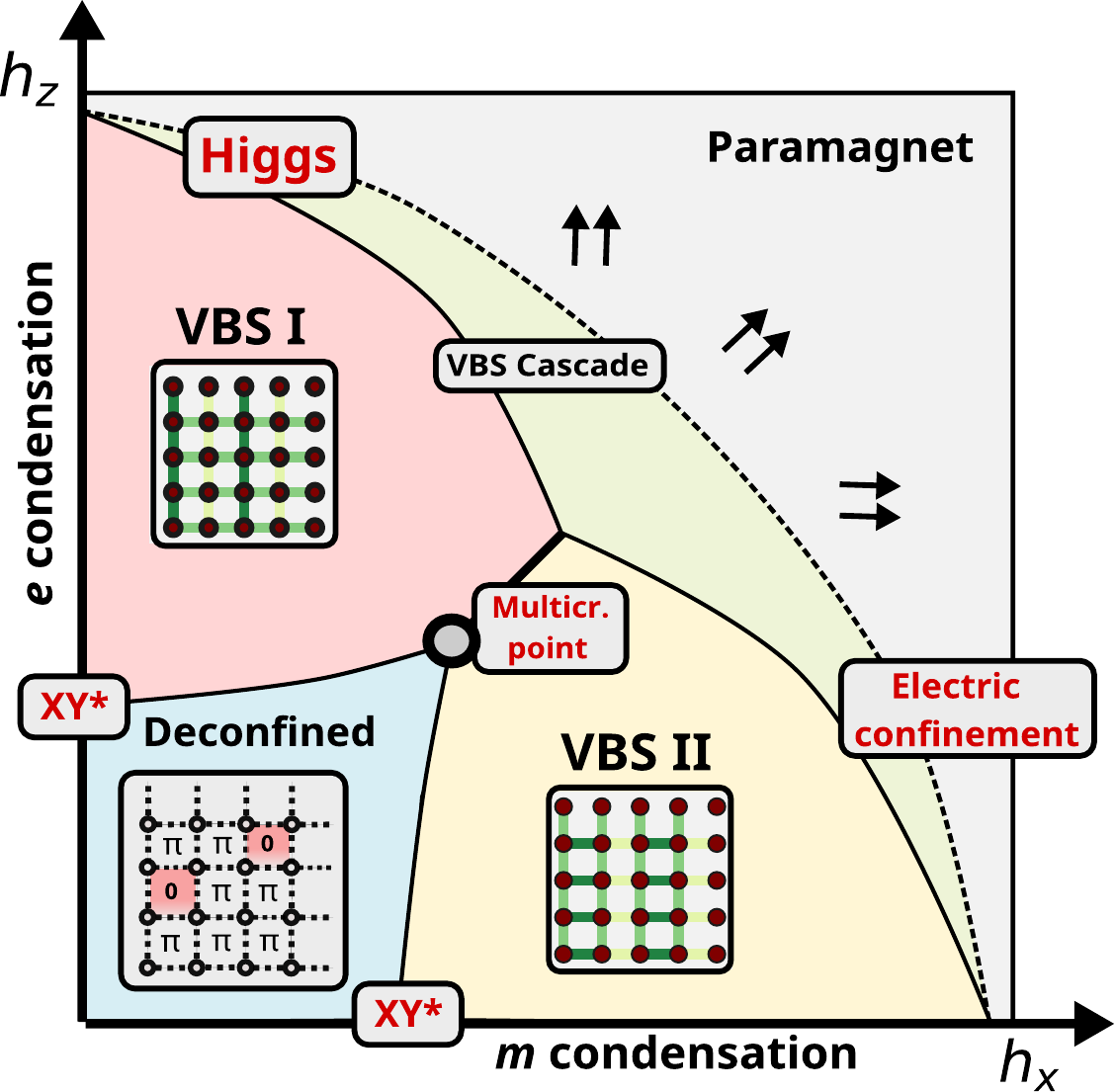}
	\caption{Quantum phase diagram of the odd Toric Code in the $h_x$ versus $h_z$ plane, see Eq. \eqref{eq:H_odd_TC}. The reflection symmetry with respect to the diagonal, a result of the electric-magnetic duality, is manifest. In the deconfined phase, the ground state consists of a background of $m$ and $e$ particles, and excitations are pairs of holes. The two VBS phases are dual to each other and are characterized by confinement and translational symmetry breaking. They meet on a segment of the self-dual line, where the transition is enriched by the spontaneous breaking of the duality symmetry. For large field values, the model undergoes a first-order transition into a trivial paramagnet.}
	\label{fig:phase_diagram}
\end{figure}

A paradigmatic example of the aforementioned systems is the $\Zt$ Fradkin--Shenker model \cite{Wegner_1971,Fradkin_1979, Fradkin2013}, capturing the dynamics of Ising matter fields whose interactions are mediated by Ising gauge fields. Modern perspectives on the understanding and classification of quantum phases of matter have shed new light on this important model, revealing the nature of its phases and associated phase transitions. Its deconfined phase is nowadays understood to exhibit topological order and anyonic excitations \cite{wilczek1990fractional}. Notably, it coincides in a certain limit with Kitaev's Toric Code model, a prime example of a topologically protected quantum error correction code \cite{KITAEV2003}. More broadly, related models with Ising gauge fields coupled to fermionic or bosonic matter have been shown to exhibit exotic phenomena, including deconfined criticality \cite{Assaad_2016,gazit2017,gazit2018, borla2022, Borla_2024, Homeier2023}, fractionalized Fermi liquids \cite{Meng_2021, Tsvelik_2020, Gazit_2020},  and unconventional boundary criticality \cite{verresen2022higgs}.

The breakdown of topological order is typically driven by one of two mechanisms: either the condensation of matter fields, known as the Higgs transition, or by the proliferation of Ising fluxes, giving rise to confinement. While the above two phase transitions separately belong to the Ising universality, as can be seen via duality relations, the multicritical point where both critical lines meet is still only partially understood and has been the focus of intense research efforts in recent times \cite{vidal2009,tupitsyn2010, Nahum2020, xu2024critical, bonati2022, ringel2024,Nahum_2024}. 

A gauge theory is not only defined by its matter and gauge field content, but also by the specific choice of a Gauss' law constraint, which, in turn, determines the static background charge configuration, also known as a superselection sector in the high-energy physics literature \cite{Wigner_1970,Goldstein_2024}. In the context of condensed matter physics, where gauge theories often emerge as effective low-energy descriptions of an underlying microscopic model, different choices can be physically motivated and correspond to different static backgrounds \cite{sachdev_2023}.

For a translationally invariant $\Zt$ gauge theory, there are two such choices; (i) ``even'' gauge theory, corresponding to the absence of static charges (``vacuum"), and (ii) ``odd'' lattice gauge theory \cite{Senthil_2000,Moessner_2001}, defined by an assignment of a single Ising charge on each site. Already for the pure $\Zt$ lattice gauge theory without matter, the presence of a static background of $e$ particles drastically modifies the phase diagram. In particular, as a consequence of the semionic mutual statistics between $e$ and $m$ particles, Ising fluxes condense at finite momentum so that the confinement transition coincides with the spontaneous breaking of translational symmetry into a valence bond solid (VBS) \cite{Jalabert_1991,mila2012}. Crucially, close to the transition, the clock anisotropy associated with symmetry-related VBS patterns is irrelevant,  which, in turn, modifies the universality class of the deconfinement transition to belong to the XY$^*$, with an emergent $U(1)$ symmetry \cite{Jalabert_1991,mila2012},

Resolving physical phenomena resulting from different choices of static backgrounds for the $\Zt$ Fradkin--Shenker model, to the best of our knowledge, remains an outstanding challenge and is the main thrust of this work. Specifically, we will focus on the case where a uniform background of static $e$ charges is present and, in tandem, $\pi$ fluxes are energetically favored in the deconfined phase ground state. This model can be reduced to the standard Toric Code model with \textit{positive} plaquette and vertex couplings in a tilted field, which we dub the ``odd Toric Code'' model. In particular, the evolution along the self-dual line, where the XY$^*$ Higgs and confinement transition lines intersect, holds the promise to host unconventional critical phenomena.

In this article, we numerically establish the rich zero temperature phase diagram of the odd Toric Code model, which includes an ``odd'' deconfined phase, dual patterns of confined VBS phases with varying structures, and a trivial paramagnet. We focus our attention on the self-dual line and study in detail the coincident condensation of $e$ and $m$ particles at finite momentum. We find numerical evidence suggesting the simultaneous occurrence of Higgs-confinement and spatial symmetry breaking in what appears to be a continuous transition. Moreover, the transition is further marked by the spontaneous breaking of the $\Zt$ $e$-$m$ duality symmetry. Interestingly, the multicritical point may provide a microscopic realization of the critical Higgs-Yukawa-QED field-theory scenario proposed in Refs.~\cite{jian_2017,dumitrescu_2026}. Our results are established via large-scale numerical simulations using both tensor-network-based methods and exact diagonalization.

This article is structured as follows: in Section \ref{sec:model} we introduce a microscopic model for the ``odd" Toric Code, discuss its symmetries, and establish the phase diagram in limiting cases. In Section \ref{sec:numerics} we discuss the numerical techniques adopted for this study. In Section \ref{sec:observables}, we define the observables used to characterize the quantum phase diagram and phase transitions. In Section \ref{sec:results}, we present and discuss the numerical results, and combine them to provide a thorough description of the ground state phase diagram. Finally, in Section \ref{sec:outlook}, we summarize our findings and discuss the impact and possible extensions of the present study.

\section{The model} 
\label{sec:model}
We consider the Hamiltonian
\begin{equation}
H = J_p \sum_{\rr^*} B_{\rr^*} + J_s\sum_\rr A_\rr -h_z \sum_{\rr,\eta} \sigma^z_{\rr,\eta}-h_x \sum_{\rr,\eta} \sigma^x_{\rr,\eta},
\label{eq:H_odd_TC}
\end{equation} 
which describes the square lattice Toric Code model in the presence of a tilted external field. The degrees of freedom are Ising spins residing on lattice bonds $b={\rr,\eta}$, with $\rr$ being a lattice site and $\eta=\hat{x}/\hat{y}$ one of the primitive lattice vectors. The first two terms consist of the plaquette operator $\hat B_{\rr^*}=\prod_{b\in \square_{\rr^*}} \hat\sigma^z_b$, where the index $\rr^*$ labels the sites of a dual lattice formed by the centers of the plaquettes, and the star operator $\hat A_{\rr}=\prod_{b \in +_\rr}\hat\sigma^x_{b}$, i.e. the product of $\hat\sigma^x_b$ on the four links emanating from the vertex $\rr$. 

We will focus on the case where both $J_s$ and $J_p$ are positive. Unless otherwise specified, we consider the symmetric case $J_s=J_p=J$, and measure all energy scales in units of $J$. In the following, we will refer to the $h \rightarrow 0$ limit of Hamiltonian \eqref{eq:H_odd_TC} as the ``odd Toric Code'', while we will call ``even Toric Code'' its more familiar counterpart with negative signs in front of the star and plaquette operators.

The external field $\sigma^z$ induces matter fluctuations, since acting with $\sigma^z$ on a bond results in flipping two neighboring star operators, nucleating two Ising charges across the bond. Similarly, $\sigma^x$ plays the role of an electric field, which creates a pair of Ising fluxes, and it is therefore responsible for magnetic fluctuations. Although, at first sight, the flipped signs of the vertex and plaquette couplings may seem like a minor modification, this choice enriches the ground state phase diagram, introducing intriguing exotic features that we unveil through a scrupulous numerical study. 

\subsection{Dual gauge theory description}
From a broader perspective, using the language of lattice gauge theories, our model \cref{eq:H_odd_TC} can be exactly mapped to a $\Zt$ lattice gauge theory coupled to Ising matter described by the Hamiltonian
\begin{equation}
H = -J_s\sum_\rr \tau^x_\rr +J_p\sum_{\rr^*} B_{\rr^*} - h_z \sum_{\rr,\eta} \tau^z_\rr \sigma^z_{\rr,\eta} \tau^z_{\rr+\eta} - h_x \sum_{\rr,\eta} \sigma^x_{\rr,\eta}.
\label{eq:2DFS}
\end{equation}
Here, the new set of Pauli matrices $\tau$ represents matter fields defined on the sites of the lattice. The ``even'' Gauss law case $\tau^x_{\rr}=A_{\rr}$ together with negative plaquette coupling, corresponding to a vacuum of quasiparticles without static $\Zt$ charges or Ising fluxes, was studied originally in the seminal paper by Fradkin and Shenker \cite{Fradkin_1979}, and more recently in \cite{vidal2009,tupitsyn2010, Nahum2020, bonati2022,xu2024critical, ringel2024}. 

By contrast, \cref{eq:H_odd_TC} is obtained after eliminating the Ising matter fields $\tau$ by resolving the ``odd'' Gauss law 
\begin{equation}
    \tau^x_{\rr}=- A_{\rr}
\end{equation}
and fixing the ``unitary'' gauge $\tau^z=+1$. The negative sign in Gauss's law carries the physical interpretation of a uniform distribution of static $\Zt$ charges, acting as fixed sources of electric lines. Similarly, the flux term is minimized by a $\pi$-flux pattern across the lattice. It is worth pointing out that the models \eqref{eq:H_odd_TC} and \eqref{eq:2DFS} exhibit some important differences in the presence of physical boundaries \cite{verresen2022higgs}, which in this paper we will not investigate.

\begin{figure}[t]
\captionsetup[subfigure]{labelformat=empty}
    \subfloat[\label{subfig:sp_dual}]{}
    \subfloat[\label{subfig:xz_dual}]{}
    \subfloat[\label{subfig:sd_vbs_configs}]{}
    \subfloat[\label{subfig:sd_vbs_configs_2}]{}
    \hspace{-10pt}
	\includegraphics[width=\linewidth]{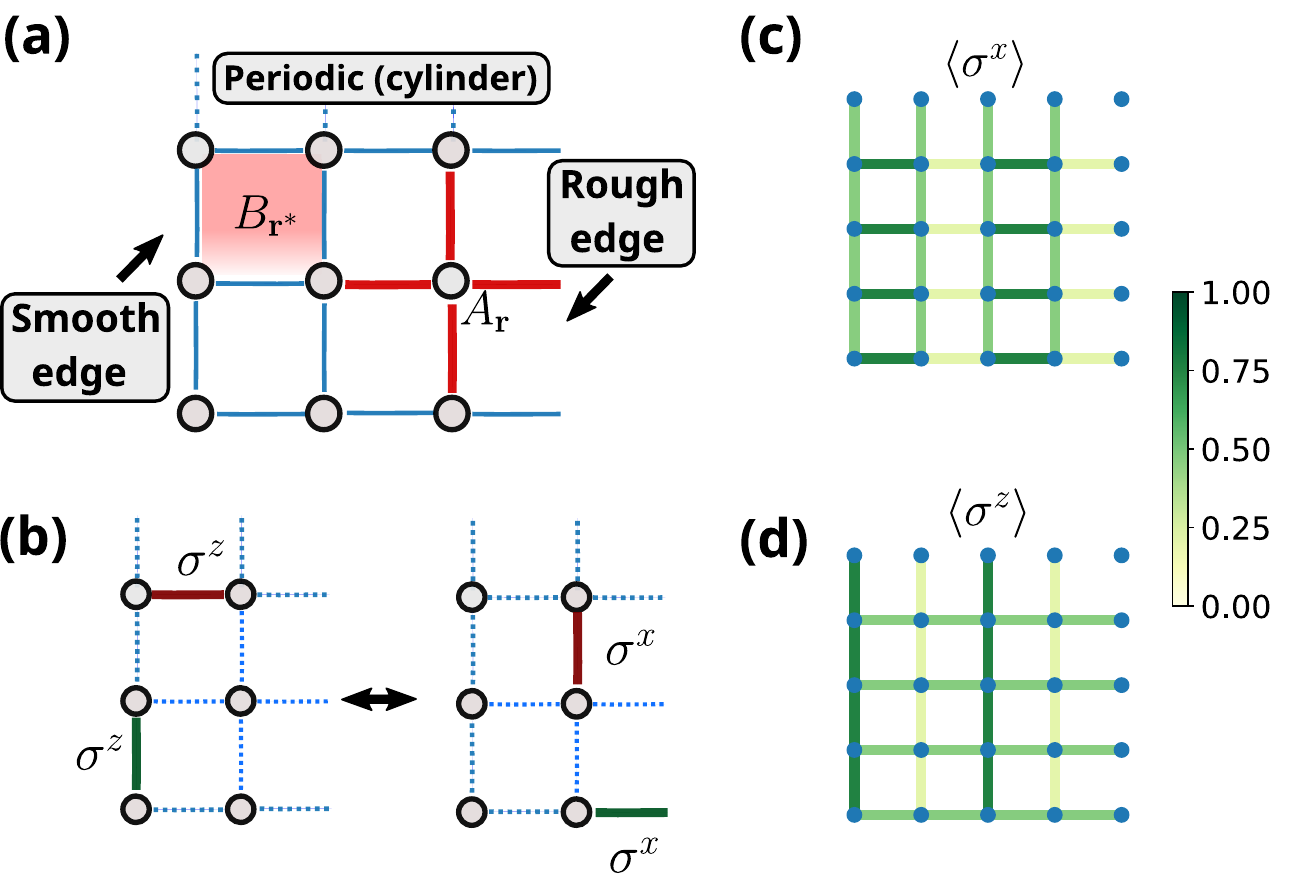}
	\caption{(a) Illustration of a 2x2 lattice geometry. The $y$ direction is taken to be periodic. For the $x$ direction, we consider open boundary conditions with a ``smooth'' left edge and a  ``rough'' right edge. This choice allows each plaquette $B_{\rr^*}$ to have a dual star operator $A_{\rr}$, which, by convention, we place at its bottom right, as shown in the figure. (b) Examples of electric-magnetic duality mappings between $\sigma^z$ and $\sigma^x$ operators on different links. The mapping is a combination of $\mathcal{U}_{em}$ transformations \eqref{eq:sd} and horizontal reflections $\mathcal{R}_h$. Following this dictionary, plaquettes are mapped to stars, and the two VBS patterns in panels (c) and (d) are mapped into each other. Expectation values of (c) $\sigma^x$ and (d) $\sigma^z$ on each link deep in the VBS phases along the lines $h_z=0$ and $h_x=0$, respectively. The two symmetry-breaking patterns are related by the electric-magnetic duality.}
	\label{fig:SD}
\end{figure}

For most of the paper we will focus on the Hamiltonian in \cref{eq:H_odd_TC}, which we also use for all the numerical simulations. It will be occasionally instructive, however, to refer to the Ising gauge theory language of \cref{eq:2DFS}, which, in certain cases, allows for a more physical interpretation of the results.

\subsection{Symmetries}
\label{ssec:symmetries}
In the absence of the external field $h_x$ ($h_z$) the model is invariant under magnetic (electric) one-form symmetries \cite{gaiotto2015generalized, mcgreevy2023generalized} generated by Wilson ('t Hooft) loops, defined as 
\begin{equation}
    W^e_{\mathcal{C}} = \prod_{\mathcal{C}} \sigma^z \qquad \qquad W^m_{\Gamma} = \prod_{\Gamma} \sigma^x
    \label{eq:loops}
\end{equation}
where $\mathcal{C}$ and $\Gamma$ are closed contours formed by consecutive links belonging to the direct and dual square lattice, respectively. See \cref{app:FM_def} for further details.

In addition, under a discrete version of the electric magnetic duality, the Hamiltonian \eqref{eq:H_odd_TC} transforms to itself after redefining the couplings as $h_x\leftrightarrow h_z$. The duality mapping exchanges the roles of the star and plaquette operators. The links (identified by their midpoints) of the original lattice are mapped to the links of the dual lattice formed by the centers of the plaquettes. The transformation can be written explicitly as
\begin{equation}
    \mathcal{U}_{em}:\sigma^z_{\rr,\rr+\hat{x}}\rightarrow \sigma^x_{\rr+\hat{x},\rr+\hat{x}-\hat{y}}, \qquad \sigma^z_{\rr,\rr+\hat{y}}\rightarrow \sigma^x_{\rr,\rr+\hat{x}}.
    \label{eq:sd}
\end{equation}
Equivalently, this can be seen as a diagonal translation of half a unit cell, which maps the sites to the centers of the plaquettes ($\rr \rightarrow \rr^*$) and horizontal links to vertical ones. 

In our finite-size numerical simulations, we will consider open and periodic boundary conditions along the $x$ and $y$ axis, respectively. More specifically, the left (right) boundary is ``smooth" (``rough") terminating in a column of plaquette (vertex) operators, as shown in \cref{subfig:sp_dual}. Although the breaking of translational symmetry invalidates the duality symmetry \eqref{eq:sd}, the Hamiltonian remains invariant under a combination of \eqref{eq:sd} and horizontal reflections $\mathcal{R}_h$ with respect to a vertical axis positioned at $x=L/2$. To see this, we note that the operators living on horizontal links of the right ``rough'' edge have no dual under \eqref{eq:sd}, but they can be mapped instead to the ones on the left ``smooth edge''. The duality exchanges the roles of the left and right edges together with the role of Ising fluxes and charges. Examples of the mapping are shown in Fig. \ref{subfig:xz_dual}. For the rest of the paper, we will refer to the enhanced duality simply as the electric-magnetic duality.
Along the self-dual line $h_x=h_z=h$, the duality becomes an exact $\Zt$ symmetry.  
\begin{figure*}[t]
\captionsetup[subfigure]{labelformat=empty}
    \subfloat[\label{subfig:density_plot_fm}]{}
    \subfloat[\label{subfig:density_plot_vbs}]{}
    \subfloat[\label{subfig:density_plot_flux}]{}
	\includegraphics[width=\linewidth]{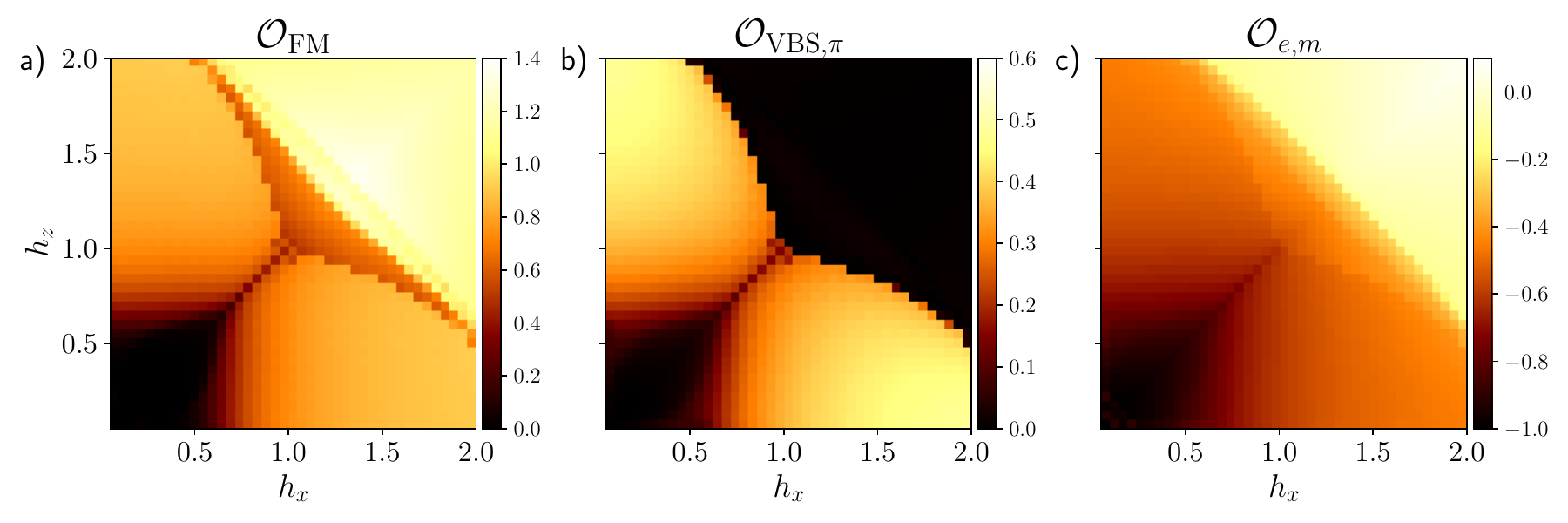}
	\caption{Density plots of the (a) FM order parameter, (b) VBS order parameter, and (c) the maximum between the average Ising flux and $\Zt$ charge, across the entire $h_x$ versus $h_z$ plane for a finite cylinder of circumference $L_y=8$. Vanishing values of the FM order parameter provide a clear-cut indicator of the deconfined topologically ordered phase, residing in the low $h_x$ and $h_z$ region. Similarly, the VBS order parameter effectively captures the frustration-induced translational symmetry breaking. The scan also reveals the presence of an intermediate region separating the VBS and trivial phases, exhibiting Higgs/confinement and a distinct VBS pattern, not captured by the $\pi$ modulated order parameter. Note that the plots are symmetric by construction, since we exploit the electric-magnetic duality of the model and only perform DMRG scans of the bottom half region $h_x \geq h_z$. }
	\label{fig:density_plots}
\end{figure*}

\subsection{Limiting cases} \label{ssec:lim_cases}
The salient features of the quantum phase diagram can be inferred by considering several limiting cases. Taking $h_x=h_z=0$ gives rise to an ``odd'' Toric Code where the star and plaquette couplings are flipped relative to the more conventional ``even'' case. Since the Hamiltonian is a sum of commuting projectors \cite{KITAEV2003}, the ground state can be constructed analytically in analogy with the even case. There, the wave function is characterized by the absence of charges ($e$-particles) and Ising fluxes ($m$-particles). In the odd case the opposite is true, and we can write
\begin{equation}
    |0\rangle_{odd} = |n^e_{\rr}=1, n^m_{\rr^*}=1\rangle \qquad \forall \rr , \rr^{*}
\end{equation}
where $n^e$ and $n^m$ are the ``number operators'' for $e$ and $m$ particles respectively. The condition $A_\rr=-1,\, \forall \rr$ is satisfied by all states where an odd number of electric lines (i.e. lines where $\sigma^x=-1$) sticks out of each vertex. The condition $B_{\rr^*}=-1$ is then satisfied by an equal weight (in absolute value) superposition of such states, where terms related by the application of a single plaquette operator carry opposite sign. An explicit construction of the ground state is provided in \cref{sec:GS_wf}. Similarly to its even counterpart, this ground state exhibits deconfinement and topological order, with a four-fold degenerate ground state on the torus. Negative star and plaquette operators are to be interpreted as $e$ and $m$ particles, respectively, which in the ground state form a uniform background. Small $h_z$ ($h_x$) external fields introduce electric (magnetic) fluctuations, creating pairs of holes in the $e$ ($m$) background. This is to be contrasted with the even Toric Code, where excitations are pairs of particles on top of the ``empty'' state.

In the opposite limit $h_x$ or $h_z \rightarrow \infty$ spins are polarized along the $\hat x$ and $\hat z$ axis, respectively, leading to a trivial quantum paramagnet. These two strong coupling regimes can be connected smoothly without going through any phase transition by rotating the spins in the $x-z$ plane. The fact that only a single phase of matter occurs in this regime is less obvious in the language of equation \eqref{eq:2DFS}, where the limits $h_x\rightarrow \infty$ and $h_z\rightarrow \infty$ have two very different physical interpretations. The former corresponds to tight confinement of the dynamical Ising matter fields ($e$-particles), while the latter is a Higgs phase characterized by $\Zt$ charge condensation.

Finally, in the absence of ``matter coupling'' $h_z=0$, the model can be seen as a pure odd $\Zt$ LGT where the Gauss law $A_\rr = -1$ is imposed energetically as long as $h_x\ll 1$. Having already established a deconfined phase for small values of $h_x$ and a paramagnetic phase as $h_x\rightarrow \infty$, we now determine the phase diagram at intermediate $h_x$ values. Progressively increasing $h_x$ will lead to confinement, resulting in a positive expectation value for the electric field $\sigma^x$. However, for moderate values of the electric field coupling, the competition with the energetic constraint $A_\rr=-1$ generates frustration, preventing the formation of a trivial uniform phase with $\sigma_b^x=1$ on all bonds. Minimizing the energy of electric lines under this constraint gives, on each site, three bonds with $\sigma_b^x=1$ and one with $\sigma^x_b=-1$. A natural candidate for such a state is then a VBS, consisting of length one electric lines connecting neighboring sites, as shown in \cref{subfig:sd_vbs_configs} and \cref{subfig:sd_vbs_configs_2}. Due to the large degeneracy of configurations satisfying the above condition, the resulting ground state pattern sensitively depends on quantum fluctuations and can be determined only through a numerical study. From a complementary perspective, in the vicinity of the deconfined-VBS transition, the low-energy physics is understood in terms of $m$ particles hopping in the background of a $\pi$-flux lattice, generated by the static background of $e$ particles. The magnetic field term $h_x$ controls the kinetic energy, such that for sufficiently large $h_x$, $m$ particles condense. The field theoretical description \cite{Jalabert_1991,sachdev_2023} involves a single complex scalar field with $D_8$ symmetry, corresponding to two degenerate minima of the $m$ particle dispersion at momenta $(0,0)$ and $(0,\pi)$. The anisotropy turns irrelevant near criticality, giving rise to an emergent $U(1)$ symmetry, such that the transition belongs to the $XY^*$ Wilson-Fisher universality. The case of $e$ particle condensation for $h_x=0$ and finite $h_z$ follows the same physical description via the aforementioned electric-magnetic duality.

From the limiting cases described here, the phase diagram displayed in Fig. \ref{fig:phase_diagram} emerges. The reflection symmetry with respect to the diagonal, due to self-duality, is manifest. While the qualitative nature of the four extended regions is undisputed, far less clear is what happens in the central region, especially in the vicinity of the self-dual line, where all four phases can potentially meet. A number of scenarios are possible, which we will attempt to clarify through a careful numerical analysis. In particular, the fate of the multicritical point associated with the non-trivial coincidence of the two critical $U(1)$ lines corresponding to the deconfinement-VBS transitions is an outstanding question. Furthermore, the frustration induced by the restriction to the ``odd" sector can be resolved energetically by several competing states. This may potentially give rise to more general VBS patterns, as we demonstrate below.

\section{Numerical methods}
\label{sec:numerics}
The positive star and plaquette terms in the Hamiltonian \eqref{eq:H_odd_TC} induce frustration, introducing a non-trivial sign structure which hinders the use of standard Monte Carlo techniques that were successfully used to study the ``even'' counterpart of the problem \cite{tupitsyn2010}. To circumvent the problem, we rely on the Density Matrix Renormalization Group (DMRG) and exact diagonalization methods.

\subsection{DMRG}
Due to the complexity of the problem, it is crucial to attempt an extrapolation of the results to the thermodynamic limit, which can be achieved through finite size scaling. In the context of matrix product states (MPS) based DMRG, this is best obtained by studying finite systems with aspect ratio near one. As explained above, we choose periodic boundary conditions in the y-direction and open in the x-direction, with a smooth left boundary and a ``rough'' right boundary, so that the number of complete stars and plaquettes is the same. This choice breaks reflections, but it is crucial to avoid explicitly breaking the self-duality symmetry. We consider systems with linear dimension $L=4,6,8,10$, with the latter being at the very limit of what is achievable with state-of-the-art computational techniques. We provide details on the numerical implementation and benchmarking in \cref{app:numerics}.

\subsection{Exact diagonalization}
Besides providing a useful benchmark for smaller systems, exact diagonalization can be used to extract exact information about the low-lying energy levels. State-of-the-art methods allow us to compute the first few eigenstates on a $4\times 4$ torus for a given momentum sector, exploiting the translational invariance of the system in both directions. We use the numerical library \texttt{QuSpin} \cite{QuSpin}, which allows us to encode basis states of spin $1/2$ systems as bit strings, and express their symmetries in terms of efficient bitwise operations. For a given momentum sector $(k_x, k_y)$, the Hilbert space of $32$ spins has a dimension of order $2^{28}$.

\section{Observables}
\label{sec:observables}
To study the quantum phase diagram of the model and identify the location and nature of the phase transitions, we rely on different observables that can be readily computed once the ground state of the system is obtained as a finite MPS with the DMRG algorithm. 

\subsection{The Fredenhagen-Marcu order parameter}
To detect deconfinement transitions, we compute the Fredenhagen-Marcu (FM) order parameter \cite{Fredenhagen_1986, Gregor_2011}, which circumvents the charge screening problem appearing in the standard Wilson-loops-based approach. This observable has been successfully employed in recent works \cite{Verresen_2021, Borla_2024, xu2024critical, linsel2025independentemanyonconfinement}, becoming increasingly popular thanks to the development of accurate numerical methods. 

The Wilson and 't Hooft loops $W^e$ and $W^m$ are defined in \cref{eq:loops}. Analogously, one can define Wilson and 't Hooft lines $\tilde{W}^e_{\tilde{\mathcal{C}}}$ and $\tilde{W}^m_{\tilde{\Gamma}}$ along open contours $\tilde{\mathcal{C}}$ and $\tilde{\Gamma}$. Note that in the gauge theory language of \cref{eq:2DFS} lines of $\sigma^z$ operators are not gauge invariant at their endpoints, and need to be terminated with Ising matter fields $\tau^z$. Since the Toric Code model in \cref{eq:H_odd_TC} can be obtained by imposing the gauge-fixing condition $\tau^z=1$, the endpoints are pinned, and open Wilson lines can be defined without modifications. Long-range order in these string operators can then be used to detect condensation of $e$ and $m$ particles, respectively. To obtain a proper order parameter, one defines the ratio

\begin{equation}
    \mathcal{O}^e_{\text{FM}} = \frac{\tilde{W}^e_{\text{half}}}{\sqrt{W^e_{\text{square}}}}
    \label{eq:FM_def},
\end{equation}

where the numerator is the expectation value of a Wilson line extending over half a loop and the denominator is the square root of the full Wilson loop. Analogously, one can define the magnetic Fredenhagen-Marcu order parameter by using half and full 't Hooft loops instead.

We also note that in the presence of translational symmetry breaking, the FM order parameter develops a spatial dependence and is affected by the boundary conditions. For this reason, we consider its maximal value within a unit cell, in the bulk, see \cref{app:FM_def} for details.

\subsection{VBS order parameter}

Next, we devise means to detect VBS patterns and, in particular, the anticipated columnar structure exhibits modulations at wave vector $\pi$ oriented along the $x$ or $y$ axis. For cylindrical geometry, preserving translation along the $y$ axis motivates the use of a Fourier transform along $y$. We define 
\begin{align*}
    \mathcal{S}_{VBS}^{X,\alpha}(x, k_y) &= \sum_{y} \langle \sigma^x_\alpha(x, y)  \rangle e^{i y \, k_y}
\end{align*}
and the equivalent $\mathcal{S}^{Z,\alpha}_{\rm VBS}(x, k_y)$ for $\sigma^z$, where the index $\alpha\in\{h,v\}$ distinguishes the horizontal (along $\hat{x}$) and vertical (along $\hat{y}$) bonds emanating from a given site and the momentum is quantized according to $k_y \in \{2\pi m /L_y\}$. 

Given the open boundary conditions in the $x$ direction, we define the order parameter 

\begin{equation}
    \mathcal{O}^{X,\alpha}_{\text{VBS}}(k_y) = \frac{1}{L^2} \sum_{x} (-1)^{x} \mathcal{S}^{X,\alpha}_{\text{VBS}}(x,k_y).
    \label{eq:ssb_op_G}
\end{equation}
which is designed to detect a columnar structure with $\pi$  modulations in the non-periodic direction. An equivalent definition holds for $\mathcal{O}_{\text{VBS}}^{Z,\alpha}$.
In particular, as explained in section \ref{ssec:lim_cases}, we anticipate that for $h_x>h_z$ the lowest energy VBS pattern corresponds to columnar order, corresponding to $(\pi,0)$ modulations which are detected by 
\begin{equation}
    \mathcal{O}^{X}_{\text{VBS,}\pi} := \mathcal{O}^{X, h}_{\text{VBS}}(k_y=0).
\end{equation}
The dual pattern, corresponding to columnar order on the dual lattice, is detected by 
\begin{equation}
\mathcal{O}^{Z}_{\text{VBS,}\pi}  := \mathcal{O}^{Z,v}_{\text{VBS}}(k_y=0).    
\end{equation}

Due to the specific choice of boundary termination at the left and right edges (see Fig. \ref{subfig:sp_dual}), particular attention must be paid to the following subtleties in order to mitigate boundary effects. When computing $\mathcal{O}_{\text{VBS}}^{Z,\alpha}$, we exclude the first column from the computation ($x\neq 1$). This is justified since star operators are absent, and hence, $\sigma^z$ operators along this column are subject to different constraints compared to the bulk. For the same reason, we exclude the last column ($x\neq L_x$) when computing $\mathcal{O}_{\text{VBS}}^{X,\alpha}$.

\subsection{Self-duality breaking}
\label{ssec:sd_break_op}
Along the self-dual line, we can test the possibility of a spontaneous breakdown of the $\Zt$ self-duality symmetry. This intriguing phenomenon was observed in the even Toric Code model \cite{Nahum2020, xu2024critical}, and in the odd case studied here a potential interplay with spatial symmetry breaking could lead to even more exotic features. Given that our choice of boundary conditions preserves the duality (combined with reflection), as explained in \cref{ssec:symmetries}, the simplest way to infer whether this symmetry is broken spontaneously is to study the difference between the expectation values of star and plaquette operators averaged over the whole lattice.  However, strict spontaneous symmetry breaking can only be observed in the thermodynamic limit. Since we work with a finite system, we introduce a symmetry-breaking parameter $\delta $, introduced by modifying the star and plaquette coefficients in opposite directions
\begin{equation}
\label{eq:deltaJ}
    J_p\rightarrow J(1+\delta), \qquad \qquad J_s\rightarrow J/(1+\delta).
\end{equation}

We then compute the linear response of the order parameter to a small change $\delta$:
\begin{equation}
    \mathcal{O}_{SD}(\delta ) = \frac{1}{L^2\, \delta}{}\sum_{\mathbf{r}} \langle A_\mathbf{r}-B_{\mathbf{r}^*} \rangle
    \label{eq:SD_break_op} 
\end{equation}

for decreasing values of the perturbation $\delta$. By simultaneously taking the limit $\delta \rightarrow 0$ and extrapolating to large values of $L$, we obtain the susceptibility associated with the perturbation induced by $\delta$, serving as a sharp order parameter for self-duality breaking.

\begin{figure}[t]
\captionsetup[subfigure]{labelformat=empty}
    \subfloat[\label{subfig:FM_hz}]{}
    \subfloat[\label{subfig:VBS_hz}]{}
    \subfloat[\label{subfig:modulations_hz}]{}
    \subfloat[\label{subfig:jumps_hz}]{}
    \hspace{-10pt}
	\includegraphics[width=\linewidth]{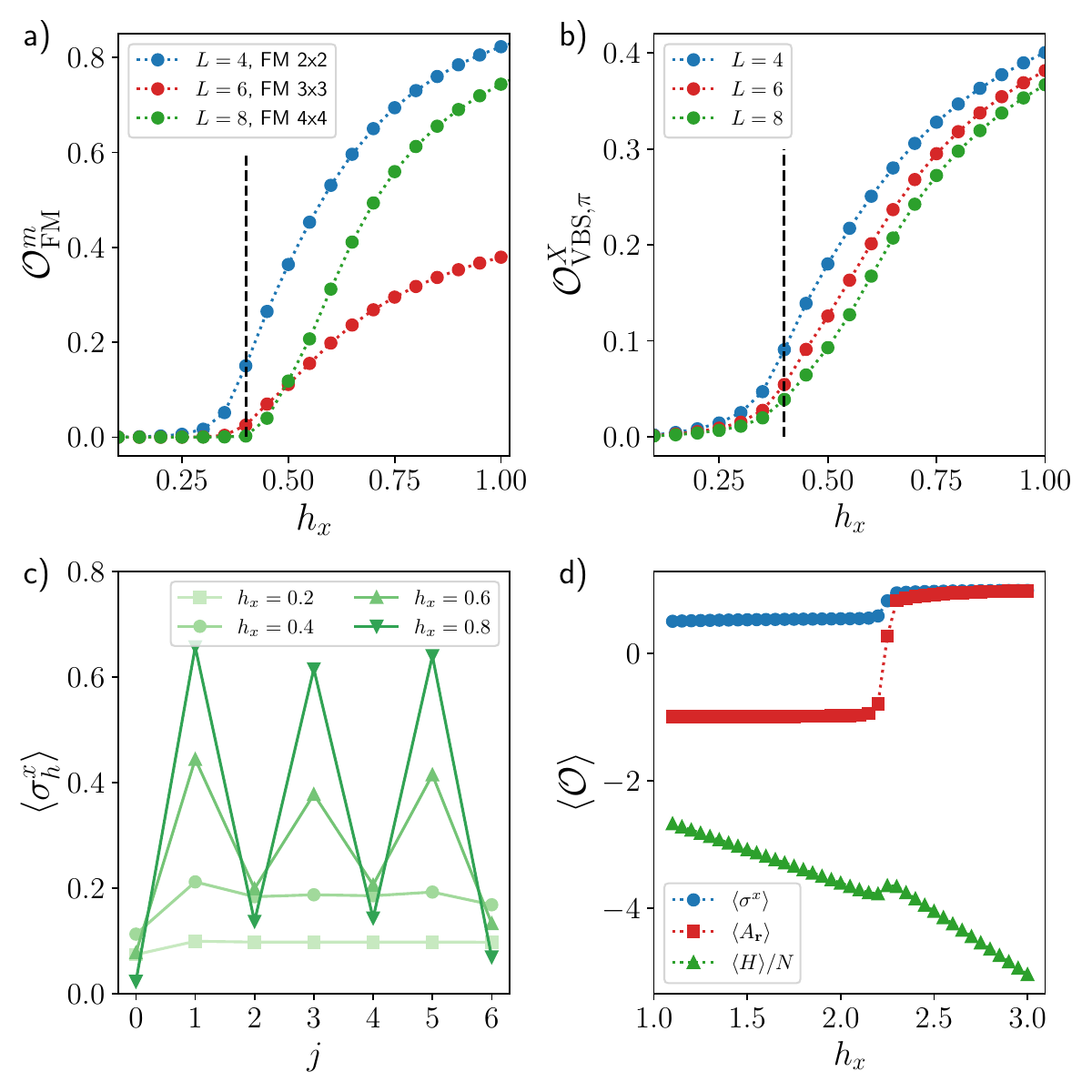}
	\caption{DMRG results for the expectation values of relevant observables away from the self-dual line, along the parameter cut $h_z=0.1$ as a function of $h_x$, for different system sizes $L$. (a)  The Fredenhagen-Marcu order (b) the columnar VBS order parameter. The vertical dashed line marks the critical field $h_{x,c}\approx 0.4$ associated with the VBS-confinement transition. (c) Profile of $\sigma^x_h$ along a cut in the horizontal (non-periodic) direction for $L=8$. Each curve corresponds to a different $h_x$ value.(d) Expectation values of energy operators $\langle A \rangle$, $\langle \sigma^x \rangle$ and $\langle H \rangle$.}
	\label{fig:hz_OPs}
\end{figure}

\subsection{Location of the critical points}
\label{subsec:location}
\begin{figure*}[t]
\captionsetup[subfigure]{labelformat=empty}
    \subfloat[\label{subfig:FM_SD}]{}
    \subfloat[\label{subfig:VBS_SD}]{}
    \subfloat[\label{subfig:SDB_SD}]{}
    \hspace{-10pt}
	\includegraphics[width=\linewidth]{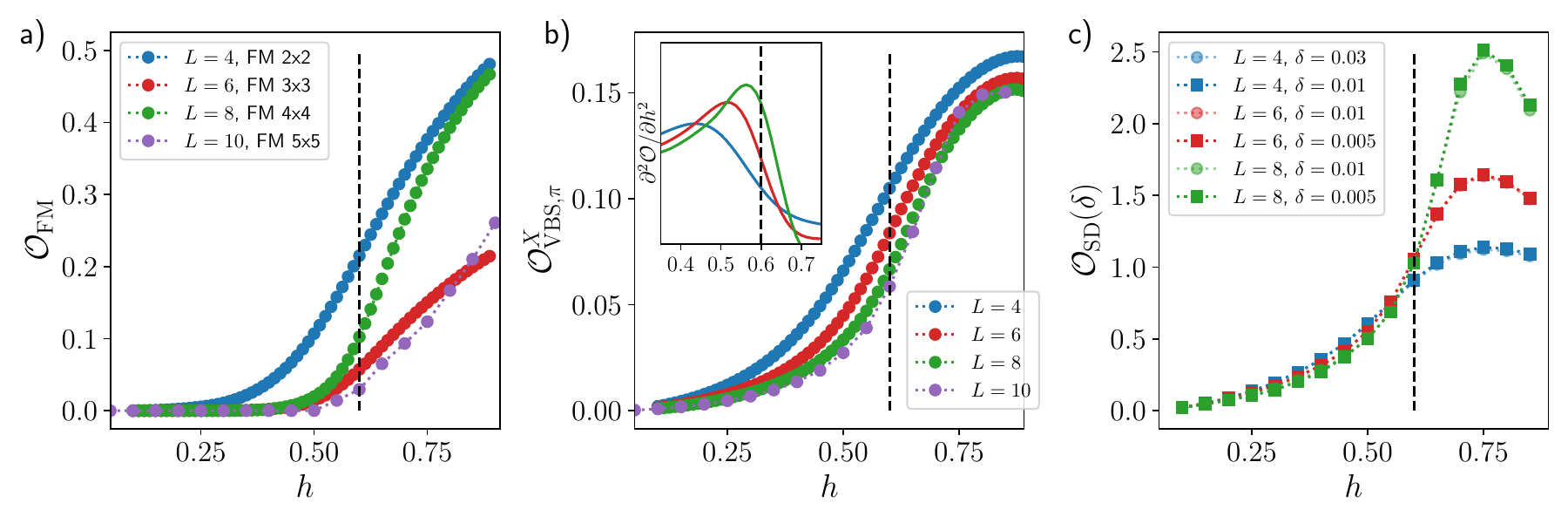}
	\caption{ DMRG results for the expectation values of relevant observables along the self-dual line $h_x=h_z=h$, for different system sizes $L$: (a)  Fredenhagen-Marcu order parameter (b) Columnar VBS order parameters (c) Self-duality breaking susceptibility. In (c) for each value of $L$ we present curves for two separate values of the symmetry-breaking perturbation $\delta$, to demonstrate that the simulations are in the linear response regime. 
    Vertical dashed line marks the numerical estimate of the multicritical point $h_c=0.6(1)$. } 
	\label{fig:SD_OPs}
\end{figure*}

While the order parameters defined above are designed to distinguish the different possible regions in the phase diagram of \cref{eq:H_odd_TC}, detecting the phase boundaries is only possible through finite-size scaling. Due to technical limitations, we can only perform simulations up to system sizes $L=8$ or $10$, with the latter becoming extremely challenging when the system becomes gapless in the vicinity of quantum critical points (QCP). For the FM order parameter, the sharpening as a function of $L$ is, in practice, sufficient to make an estimate of the QCP. For the VBS order parameter, the sharpening is less visible, and to pinpoint the QCP, we look at the position of the diverging with system size peaks in its second derivative, as shown, for example, in \cref{subfig:VBS_SD}. Lastly, in the thermodynamic limit, the self-duality breaking susceptibility \cref{eq:SD_break_op} is expected to be finite in the deconfined phase and diverge in the confined phase. The QCP is then signaled by the crossing point of finite-size curves.

\section{Results}
\label{sec:results}
In the following section, we present numerical results aimed at mapping out the quantum phase diagram of Hamiltonian \eqref{eq:H_odd_TC}, based on the observables defined above. Our first line of attack will be to determine the phase diagram topology in the $h_x$-$h_z$ plane, followed by a detailed investigation of the self-dual line.

\subsection{Global phase diagram in $h_x$-$h_z$ plane}
We begin our analysis by carrying out DMRG scans of the phase diagram in the $h_x$-$h_z$ plane, for $L_y=8$. To map out the phase boundaries, we primarily focus on the Fredenhagen-Marcu and VBS order parameters. The former is non-vanishing everywhere outside of the deconfined phase, while the latter is expected to detect the modulations in the ground state wave function and attain a finite value exclusively in the VBS phases. 
Moreover, we compute the magnetic flux and electric charge per site, averaged over the entire lattice, as captured by $\left\langle B\right\rangle=\frac{1}{L^2}\sum_{\\r*}\left\langle B_{\rr*}\right\rangle$ and $\left\langle A\right\rangle=\frac{1}{L^2}\sum_{\rr}\left\langle A_{\rr}\right\rangle$, respectively. Both quantities tend to -1, deep in the deconfined phase, and saturate for large field values deep in the paramagnetic phase to $\left\langle A\right\rangle = \left(h_x /\sqrt{h_x^2 + h_z^2}\right)^4$, and a similar expression for $\left\langle B\right\rangle$, with $h_x\leftrightarrow h_z$. 

We exploit the electric-magnetic duality of the model and only perform scans of the bottom-right triangular region $h_z \ge h_x$, since the rest of the phase diagram is readily obtained by reflecting with respect to the self-dual line and applying the duality. In particular, we compute $\mathcal{O}_{\text{FM}}=\max(\mathcal{O}^e_{\text{FM}},\mathcal{O}^m_{\text{FM}})$, $\mathcal{O}_{\text{VBS,}\pi}=\max(\mathcal{O}^X_{\text{VBS,}\pi},\mathcal{O}^Z_{\text{VBS,}\pi})$ and $\mathcal{O}_{e,m}=\max(\left\langle A\right\rangle,\left\langle B\right\rangle)$ in \cref{subfig:density_plot_fm}, \cref{subfig:density_plot_vbs}, and \cref{subfig:density_plot_flux}, respectively, so that the symmetry with respect to the self dual line is preserved. 

As anticipated, we identify a deconfined phase at low $h_x/h_z$ values with vanishing FM and VBS order parameters. As the strength of either field is increased, the deconfined phase gives way to confined VBS regions. In particular, we observe a non-zero value for $\mathcal{O}_{\text{VBS,}\pi}$, corresponding to columnar bond order. The resulting VBS pattern follows the expected flux condensation for  $h_x>h_z$, and electric charge condensation for $h_z>h_x$. The appearance of VBS order is further accompanied by a finite non-zero FM order parameter, indicating the breakdown of topological order through confinement, Higgs transitions or both. 

For sufficiently large field strength, a trivial paramagnetic phase is recovered. Interestingly, we also observe an unexpected intermediate phase, separating the VBS and paramagnetic phases. As we demonstrate below, this phase is characterized by an intricate spatial symmetry breaking pattern. The results are summarized in \cref{fig:density_plots}.

After establishing the general phase diagram topology, in the following sections, we will investigate in greater detail the phase transitions separating the deconfined and VBS phases, away from and along the self-dual line, the nature of the intermediate phase separating the VBS and paramagnetic phases, and the low-energy excitation spectrum.

\subsection{Flux condensation induced confinement and VBS order away from the self-dual line.}
To exemplify the above results, we investigate in detail a specific parameter cut by fixing $h_z=0.1$ and increasing $h_x$ to drive the confinement transition. We note that this choice breaks the self-duality. In \cref{fig:hz_OPs}, we explore the evolution of FM and VBS order parameters as a function of $h_x$ and system size.

As expected, both order parameters rise in tandem and sharpen as the system size increases, signaling a single and continuous transition. It is worth noting that the finite-size scaling of $\mathcal{O}_{\text{FM}}$ is further complicated by even-odd effects, which induce an alternating pattern in the behavior of the curves. We attribute this effect to the fact that odd values of $L/2$ are incompatible with the periodicity of the VBS, since the loops and lines on which the order parameter is defined stretch over a non-integer number of unit cells. The specific symmetry-breaking pattern is compatible with columnar order, which is revealed by the ($\pi, 0$) modulations in the expectation value of $\sigma^x$ on horizontal links. This corresponds to having a finite density of tightly confined $e$-particles, which is reflected by negative expectation values of the star operator $A_{\rr}$ while keeping the smallest possible number of ``frustrated links'' with $\langle \sigma^x \rangle=-1$. In \cref{subfig:modulations_hz}, we provide a real space depiction of the VBS pattern, by plotting $\sigma^x_h$ along a vertical cut. We indeed find an evolution toward a columnar VBS pattern as the external field $h_x$ increases above the confinement transition.

For larger values of $h_x$ the symmetry-broken phase is destroyed, leaving room for a trivial paramagnet where $\langle \sigma^x \rangle=1$ everywhere and no $e$ particles are left in the system. The transition between the two phases is first order, as signaled by sharp jumps in the energy observables $\langle \sigma^x \rangle$ and $\langle A \rangle$, shown in \cref{subfig:jumps_hz}. By inspecting the $h_x$-$h_z$ phase diagrams in \cref{fig:density_plots}, however, we notice that the first order transition point evolves into an extended intermediate region, which occupies a progressively larger area as the self-dual line is approached. This unexpected feature will be investigated further in \cref{ssec:myst}.

\subsection{Higgs-confinement transition along the self-dual line}
We now turn to the main focus of this work, which is to investigate the evolution along the self-dual line $h_x=h_z = h$ and, in particular, the multicritical point where the confinement and Higgs transition lines meet. While the weak and strong field limits are relatively understood, the fate of intermediate couplings, where condensation of $e$ and $m$ particles compete, is an outstanding question. One must consider several possibilities for the phase diagram topology and associated phase transitions. Do the two dual VBS phases touch, or are they separated by an additional, qualitatively different intermediate region? Is self-duality broken at any point along the line? Is there one or multiple quantum critical points, or maybe a first-order line ending at a continuous multicritical point as observed for the even Toric Code \cite{vidal2009,tupitsyn2010, Nahum2020, xu2024critical, bonati2022, ringel2024}?

With the above inquiries in mind, we first investigate the breakdown of topological order as $h$ is increased. To this end, in Fig. \ref{subfig:FM_SD} we depict the FM order parameters $\mathcal{O}_{\text{FM}}$ as functions of $h$ for different system sizes $L$. We observe a sharp, yet continuous rise of $\mathcal{O}_{\text{FM}}$ at $h_c=0.6(1)$, consistent with a continuous transition.  

\begin{figure*}[t]
\captionsetup[subfigure]{labelformat=empty}
    \subfloat[\label{subfig:x_and_z_sd}]{}
    \subfloat[\label{subfig:vbs_perp}]{}
    \subfloat[\label{subfig:energy_sd}]{}
    \hspace{-10pt}
	\includegraphics[width=\linewidth]{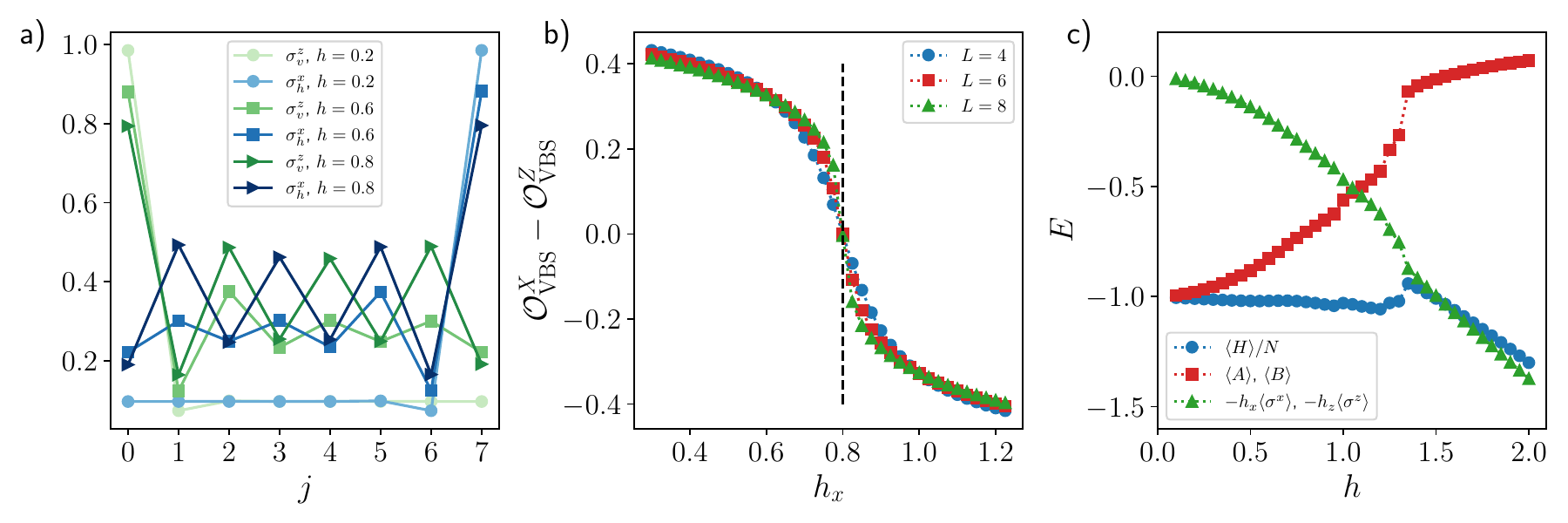}
	\caption{Expectation values of several relevant observables along the self-dual line, obtained with DMRG on a cylinder of circumference $L_y=8$. (a) Profile of $\sigma^x$ and $\sigma^z$ along a cut in the non-periodic direction, on vertical and horizontal links, respectively. The $\pi$-modulations within the VBS phase $0.6<h<1$ are evident. For a given $h$, the profiles of $\sigma^x$ and $\sigma^z$ are mirror symmetric, as follows from duality combined with reflection. 
    (b) Difference between the $\sigma^x$ and $\sigma^z$ VBS order parameters along the cut $h_z=1.6-h_x$ \textit{perpendicular} to the self-dual line, intersecting it at $h=0.8$ . This quantity can be interpreted as a duality-breaking order parameter, whose sharpening signals a first-order transition between the two different bond orderings. 
    (c) Total energy and its plaquette and field components along the self-dual line. Due to the electric-magnetic duality, the average expectation values of stars and plaquettes, and $\sigma^x$ and $\sigma^z$ are identical. 
    }
	\label{fig:vbs_corrs}
\end{figure*}

Next, we investigate if translational symmetry is broken spontaneously along the self-dual line, as it happens with the columnar VBS order observed away from the duality condition. This question is crucial for mapping out the central region of the phase diagram, as it is not guaranteed that the two VBS phases connect smoothly into each other. Indeed, the VBS order parameter, $\mathcal{O}_{\text{VBS}}$, shown in \cref{subfig:VBS_SD}, rises continuously roughly at the same value $h_c\approx0.6(1)$ computed from the deconfinement transition. To pin down $h_c$ more accurately, we numerically compute the second derivative, $d ^2\mathcal{O}^X_{\text{VBS},\pi}/d h^2$, which is expected to diverge at the critical coupling, see inset of \cref{subfig:VBS_SD}. Indeed, we identify a peak at $h_c=0.6(1)$ that diverges with system size, consistent with the above analysis. In \cref{subfig:x_and_z_sd}, we provide a real space snapshot of $\sigma^z_v$ and $\sigma^x_h$ along the $x$ axis. We observe the appearance of modulated VBS order, as the field $h$ is increased. In addition, the two VBS patterns are related by the above-mentioned $\Zt$ duality transformation, which involves reflection.

To further test translational symmetry breaking in the vicinity of the self-dual line, in \cref{subfig:vbs_perp} we consider a parameter cut perpendicular to it and crossing it at $h=0.8$. We observe an exchange between the columnar VBS patterns described by $\mathcal{O}^X_{\text{VBS, }\pi}$ and $\mathcal{O}^Z_{\text{VBS, }\pi}$ as we cross the self-dual line. With an increase in system size, the transition sharpens and appears to develop a jump discontinuity. Although it is difficult to deduce the actual thermodynamic limit using our limited finite-size scaling, the numerical data is consistent with a first-order transition separating the competing VBS patterns across the self-dual line.

Special to the self-dual line is the enlarged $\Zt$ symmetry associated with $e$-$m$ duality, which motivates exploring the exotic possibility that the self-duality symmetry is also broken spontaneously, as was found in the even case \cite{xu2024critical}. We test this scenario by computing the $\Zt$ duality breaking susceptibility $\mathcal{O}_{SD}(\delta )$ defined in \cref{eq:SD_break_op} for different system sizes across the transition. We monitored the convergence of the susceptibility in the limit $\delta \to0$. The result of this analysis is shown in Fig. \ref{subfig:SDB_SD}. Indeed, we observe a diverging response with system size for $h>h_c =0.6(1)$, consistent with the appearance of a duality-breaking phase coinciding with the deconfinement and VBS transitions; further details are given in \cref{app:duality_sus}.

Lastly, in Fig. \ref{subfig:energy_sd}, we depict the different contributions to the energy and its total value along the self-dual line. In the Toric Code limit, as expected, all stars and plaquettes contribute precisely -1. Away from this limit, as $h$ increases, the competition with the field term, which tends to polarize the spins and expel $e$ and $m$ particles from the ground state, and the other terms favoring Ising fluxes and charges, is manifest. We note that all energy terms evolve continuously across the Higgs-confinement transition, further supporting the continuous transition scenario.

\subsection{Cascade of VBS phases}
\label{ssec:myst}
Our numerical data supports the presence of a continuous transition between a deconfined phase at low external fields and a confined phase where translational symmetry is also spontaneously broken. As the external fields are further increased and the system approaches the paramagnetic phase, it is not immediately clear how translational symmetry is restored. While in the limiting cases $h_x=0$ and large $h_z$ (same for the dual case), our results are consistent with a single first-order transition. Scans of the phase diagram shown in Fig. \ref{fig:density_plots} suggest the presence of an intermediate extended phase. 

To understand its properties in more detail, we conducted refined scans that focused on the intermediate region. To that end, we consider diagonal parameter cuts $h_z=a+h_x$, for a constant shift $a$, which progressively deviates from the self-dual line as $a$ is increased. 

\begin{figure}[t]
\captionsetup[subfigure]{labelformat=empty}
    \subfloat[\label{subfig:x_myst}]{}
    \subfloat[\label{subfig:B_myst}]{}
    \subfloat[\label{subfig:xs_myst_1}]{}
    \subfloat[\label{subfig:xs_myst_2}]{}
    \hspace{-10pt}
	\includegraphics[width=\linewidth]{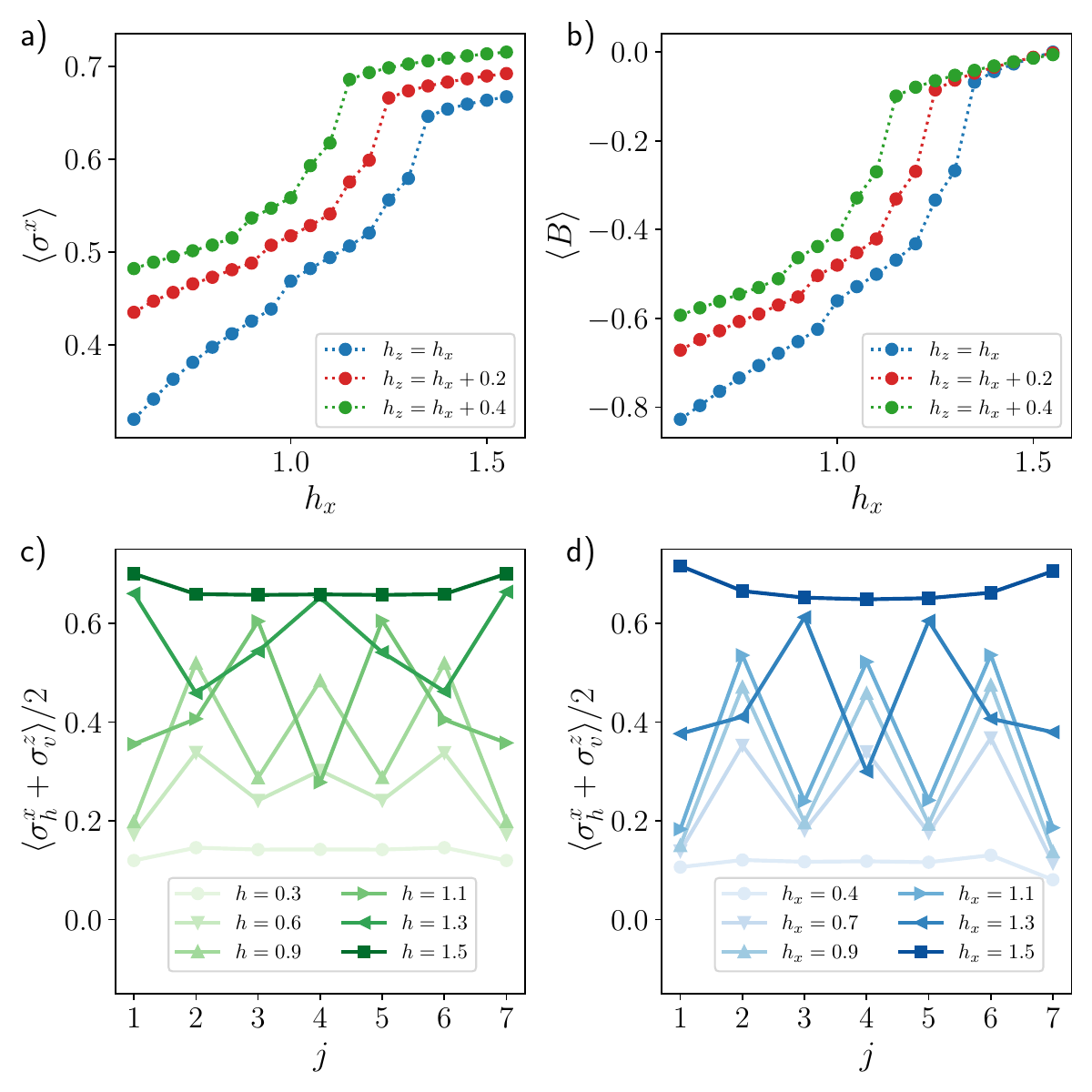}
	\caption{Average value of the (a) electric field $\sigma^x$ and (b) plaquette $\langle B \rangle$ along several parameter cuts, crossing the VBS cascade regime. Profile of the local expectation values $\langle \sigma^x_h (j) +\sigma_v^z(j) \rangle /2$, across a horizontal cut, for $L=8$, $j$ denoting the site index, (c) along and (d) parallel to the self-dual line, along the cut $h_z=h_x+0.3$. Different curves correspond to different $h_x$ values.}
	\label{fig:myst}
\end{figure}

Along these cuts, we computed the average $ \left\langle \sigma_x \right\rangle$ and flux $\left\langle B \right\rangle$, as shown Figs. \cref{subfig:x_myst} and \cref{subfig:B_myst} respectively. The evolution along the intermediate phase suggests an extended segment exhibiting a cascade of phase transitions, manifest in irregular behavior, both being characterized by steep variations interleaved by smooth behavior. 

We interpret the jumps in energy observables as a rearrangement of the bonds, distinct from the columnar VBS pattern discussed so far. To see this more clearly, Figs. \ref{subfig:xs_myst_1} and \ref{subfig:xs_myst_2}, clearly show that, for sufficiently large $h$, spatial profiles of $\frac{1}{2}\langle \sigma_x^h(j)+\sigma_z^h(j)\rangle$ along the non-periodic direction progressively deviate from $\pi$ modulations, associated with columnar order. Due to the limitations in system size, we cannot conclude whether the new pattern is incommensurate, or if it exhibits, for instance, $\pi/2$ modulations as suggested by the profile at $h=1.3$. If this is the case, one can speculate that in the thermodynamic limit, the system becomes extremely sensitive to small increases in $h$, with bond orderings characterized by progressively larger wavelengths becoming energetically favorable. 

\subsection{Low energy excitation spectrum}

\begin{figure*}[t]
    \captionsetup[subfigure]{labelformat=empty}
    \subfloat[\label{subfig:gaps}]{}
    \subfloat[\label{subfig:ED_sd}]{}
    \subfloat[\label{subfig:ED_hz}]{}
	\includegraphics[width=0.9\linewidth]{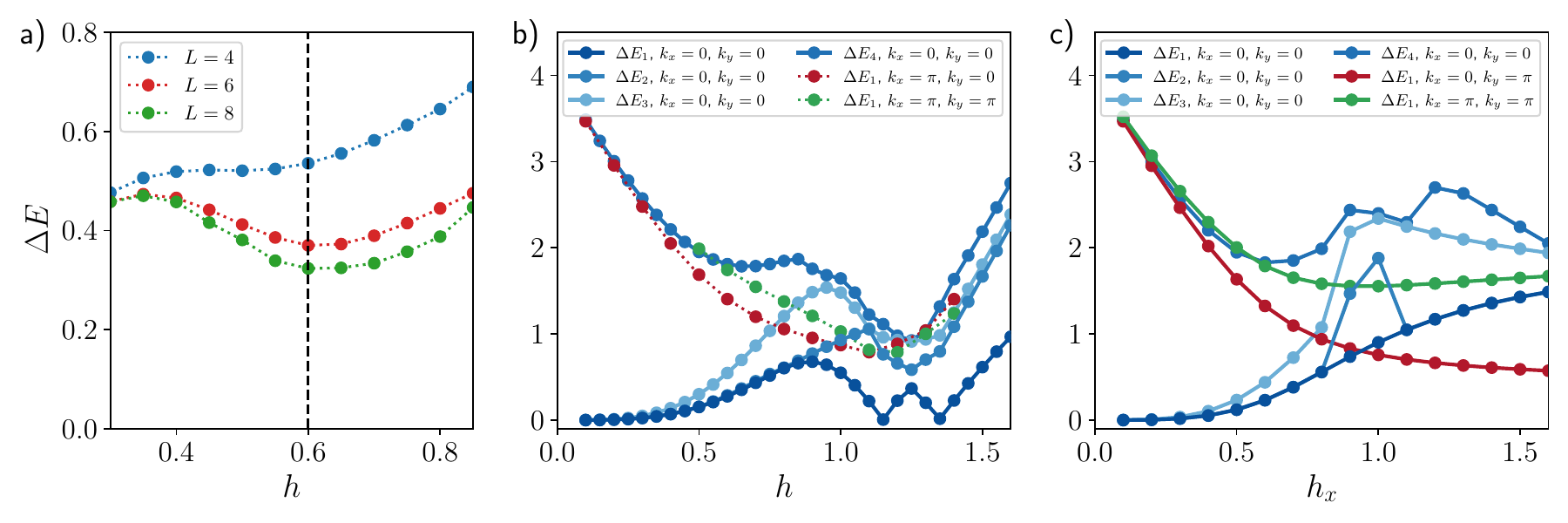}
	\caption{(a) First excitation gap along the self-dual line computed with DMRG for system sizes $L=4,6,8$ on a cylinder with open boundary conditions. The dashed line markes the position of the multicritical point $h_c=0.6(1)$. (b) and (c): low-lying energy gaps obtained by exact diagonalization for a $4\times 4$ system on a torus (b) on the self-dual line and (c) the line $h_z=0.1$. For small values of $h$, the fourfold degeneracy of the ground state is evident, and the first excitations are consistent with pairs of $e$ and $m$ particles with $\Delta E \approx 4$ in the $h\rightarrow 0$ limit. In both cases, we observe a low lying excitation at finite momentum $(\pi,0)$ near the confinement transition. }
	\label{fig:ED_4x4}
\end{figure*}

In this section, we explore the low-energy spectrum to gain further insights into the evolution of low-lying excitations across the phase diagram. Using DMRG, we are able to extract not only the ground state, but also the lowest excited state above it and its energy. It is particularly useful to compute the energy gap, whose closing upon approach to the critical point indicates a continuous phase transition. The results of this analysis along the self-dual line are shown in Fig. \ref{subfig:gaps}. We find that the energy gap softens with an increase in the system size, consistent with a quantum critical point at $h_c \sim 0.6(1)$.

As a complementary tool, we perform exact diagonalization scans for $L_x=L_y=4$ lattices defined on a torus with periodic boundary conditions. Naturally, such an analysis is limited due to the small system sizes and does not allow any extrapolation to the thermodynamic limit for this particular geometry. Nonetheless, accessing the exact spectrum of low-lying excitation provides valuable information on the nature of the transitions. 

For small values of $h$, the four-fold topological degeneracy characterizing the deconfined phase is manifest via near-degenerate low-lying states whose energy difference is consistent with an exponentially small splitting in system size. Anticipating a simultaneous condensation of $e$ and $m$ particles to drive the transition, we track the closing of the excitation gap above the four-fold degenerate ground state. At $h=0$, as expected, this takes the value $\Delta E =4$ corresponding to pairs of $e$ or $m$ excitations. Although finite size effects make it difficult to detect the precise closing of this gap, increasing $h$ continuously decreases the gap, as seen in \cref{subfig:ED_sd}. 

To better understand the physical nature of the excitations responsible for the gap closing, we resolve different momentum sectors corresponding to wave functions that are either uniform ($\mathbf{k} = (0,0)$) or exhibit $\pi$ modulations in one ($\mathbf{k} = (0,\pi)$) or both ($\mathbf{k} = (\pi,\pi)$) directions. We observe that the lowest excitation gap occurs at finite momentum ($\mathbf{k} = (0,\pi)$). As an \textit{ad hoc} definition for exiting the deconfined phase, we choose the $h$ value for which the lowest excited state energy intersects the energy of one of the ($h\rightarrow0$) topologically degenerate states, see orange and green curves in \cref{subfig:ED_sd}. This occurs at $h\approx0.7$ along the self-dual line, close to the $h_c$ found above using other methods for larger systems in a different geometry.

For larger values of $h$ along the self-dual line, beyond the Higgs-confinement transition, the gap remains small up until $h \approx 1.5$, where the system enters the paramagnetic phase. In particular, we observe two level crossings in the region $1<h<1.5$. This is consistent with the conclusion reached above, that this region is characterized by fierce competition between different symmetry-breaking patterns, resulting in multiple rearrangements of the ground state before translational invariance is fully restored. For large values of $h$ the gap increases linearly with $h$, as the excitations here are local spin flips (magnons) with $\Delta E \approx 2h$.

It is helpful to compare the ED results along the self-dual line with those along the cut $h_z=0.1$. In \cref{subfig:ED_hz}, we again find that when approaching the confinement transition, the lowest excitation carries finite momentum $(0,\pi)$.

\section{Discussion and summary } 
\label{sec:outlook}

In this work, we mapped out the intricate phase diagram of the odd $\mathbb{Z}_2$ Ising-gauge theory. Most notably, our numerical results suggest the presence of an exotic multicritical point along the self-dual line, where the two continuous XY$^*$ transitions associated with Higgs and confinement intersect. At this critical point, the simultaneous condensation of $e$ and $m$ particles is accompanied by the formation of a columnar valence bond solid and spontaneous breaking of the $\mathbb{Z}_2$ electric-magnetic duality.

A natural question is how to identify the long-wavelength field theory description and associated universal properties of the putative multicritical point. Refs.~\cite{jian_2017,dumitrescu_2026} analyzed related models enriched by a self-dual symmetry and presenting a similar phase diagram topology. The authors suggested that the multicritical point belongs to the universality class of Higgs-Yukawa-QED. The emergent $U(1)$ symmetry of the condensing $e$ and $m$ particles, combined with their non-trivial semionic mutual statistics, points toward a theory of two complex scalar fields coupled via a mutual Chern-Simons term, as explored in Ref.~\cite{Geraedts_2013}. Developing numerical probes to assess the relevance of these scenarios to our model, extract related critical data, and explore possible extensions
remains an outstanding challenge.

Recent advances in numerical methodologies may enable access to larger system sizes, approaching the thermodynamic limit. In particular, infinite projected entangled pair states (iPEPS) have been successfully applied to extract ground state properties and the FM order parameter in the even Fradkin--Shenker model \cite{xu2024critical}, suggesting that similar methods could prove valuable for studying the odd sector considered here. Approaches based on neural quantum states (NQS) \cite{Carleo_2017,Bohrdt_2024,kufel2024} also hold promise for reliable ground state optimization on lattice sizes beyond the reach of DMRG-based methods. Applying these techniques to the present model could provide a more precise characterization of the critical behavior and shed further light on the nature of the multicritical point.

Across the phase diagram, we confirmed the presence of Higgs and confinement transitions enriched by the formation of valence bond solids. Intriguingly, further increasing the field strength does not immediately lead to a trivial paramagnetic state but instead gives rise to a cascade of potentially incommensurate crystalline phases induced by frustration. We have tested this scenario numerically for moderate lattice sizes; determining its thermodynamic fate remains an open question for future work.

Lastly, it would be interesting to further generalize the Higgs-confinement multicriticality scenario in several directions. These include incorporating fermionic or bosonic matter fields, exploring gauge groups beyond $\mathbb{Z}_2$ \cite{Elio_2024,Laumann_2025}, and studying different lattice geometries \cite{linsel_2025}, where lattice-specific frustration effects may play a significant role. Furthermore, varying background charge configurations could offer additional tunability and potentially lead to qualitatively new phases and transitions. Investigating these extensions may reveal broader organizing principles for multicritical phenomena in lattice gauge theories.

\section{Acknowledgements} We acknowledge helpful discussions with Sergej Moroz, Adam Nahum, Zohar Ringel, Fabian Grusdt, Annabelle Bohrdt, Wen-Tao Xu, and Gertian Roose. We thank the developers of the tensor network libraries \texttt{TeNPy} \cite{tenpy2024} and \texttt{ITensor} \cite{ITensor}, and of the exact diagonalization library \texttt{QuSpin} \cite{QuSpin}. U.B. acknowledges support from the Israel Academy of Sciences and Humanities through the Excellence Fellowship for International Postdoctoral Researchers. U.B.~acknowledges funding by the Max Planck Society, the Deutsche Forschungsgemeinschaft (DFG, German Research Foundation) under Germany’s Excellence Strategy – EXC-2111 – 390814868, and the European Research Council (ERC) under the European Union’s Horizon Europe research and innovation program (Grant Agreement No.~101165667)—ERC Starting Grant QuSiGauge. This work is part of the Quantum Computing for High-Energy Physics (QC4HEP) working group. S.G. acknowledges support from the Israel Science Foundation (ISF) Grant no.
586/22 and the US–Israel Binational Science Foundation (BSF) Grant no.  2020264. Computational resources were provided by the Intel Labs Academic Compute Environment and the Fritz Haber Center for Molecular Dynamics, The Hebrew University of Jerusalem.

\section{Data availability} All the numerical data supporting the findings of this article are openly available \cite{Zenodo}. 

\appendix

\section{Explicit construction of the $h=0$ (``odd Toric Code'') ground state}
\label{sec:GS_wf}
\begin{figure}[t]
	\includegraphics[width=0.95\linewidth]{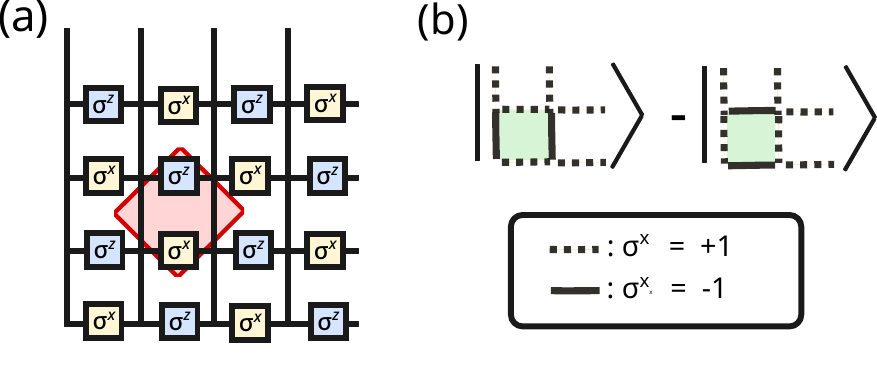}
	\caption{(a) Representation of the global transformation \eqref{eq:odd_GS} that transforms the ground state of the Toric Code into its ``odd" counterpart. In red, we show a unit cell of the dual lattice (formed by the centers of the links). (b) Example of one of the states in the superposition forming the ground state of the ``odd'' Toric Code. All the $A_{\rr}$ operators evaluate to $-1$, and under the action of $B_{\rr^*}$ on the green plaquette the state changes sign, i.e. $B_{\rr^*}$ evaluates to $-1$.}
	\label{supfig:odd_TC_GS}
\end{figure}

In the main text, we explained how the ground state of the odd Toric Code can be constructed using the fact that its Hamiltonian is a stabilizer code, i.e. a sum of commuting terms. This description is most natural in the electric ($\sigma^x$) basis, where the states in the superposition are formed by electric ``trees'' of any size, i.e. configurations where at each site a string either starts (one outgoing electric line), or it splits into two strings (one incoming and two outgoing electric lines). Such a state can be written down explicitly by taking as a reference the familiar ``even'' Toric Code state. In the $\sigma^x$ basis, this can be written as
\begin{equation}
    |0\rangle_{\textit{even}} = \prod_{\rr^*} (1+B_{\rr^*})|\uparrow \dots\rangle_x,
\end{equation}
where the operator acting on the electric vacuum creates all possible equal weight superpositions of electric loops. The global operator shown in Fig. \ref{supfig:odd_TC_GS}
\begin{equation}
    \mathcal{Q}_{TC} = \prod \sigma^x_{i,i+\hat{x}} \sigma^z_{i+\hat{x},i+2\hat{x}} \sigma^z_{i+\hat{y},i+\hat{y}+\hat{x}} \sigma^x_{i+\hat{y}+\hat{x},i+\hat{y}+2\hat{x}},
    \label{eq:odd_GS}
\end{equation}  
where the product is taken over all $2\times2$ unit cells encompassing four plaquettes, toggles between the even and odd Toric Code ground states, i.e.

\begin{equation}
    |0\rangle_{odd}=\mathcal{Q}_{TC}|0\rangle_{even}
\end{equation}
and vice versa since $\mathcal{Q}_{TC}^2=\mathbb{I}$. 

To conclude, we note that while the operator \cref{eq:odd_GS} relies on a specific choice of links which explicitly breaks the translational invariance, this choice is non-unique and it is possible to construct a ground state which does not break the translational invariance of the Hamiltonian.

\section{Details on the computation of the Fredenhagen-Marcu order parameter}
\label{app:FM_def}

\begin{figure}[t]
    \captionsetup[subfigure]{labelformat=empty}
	\includegraphics[width=0.95\linewidth]{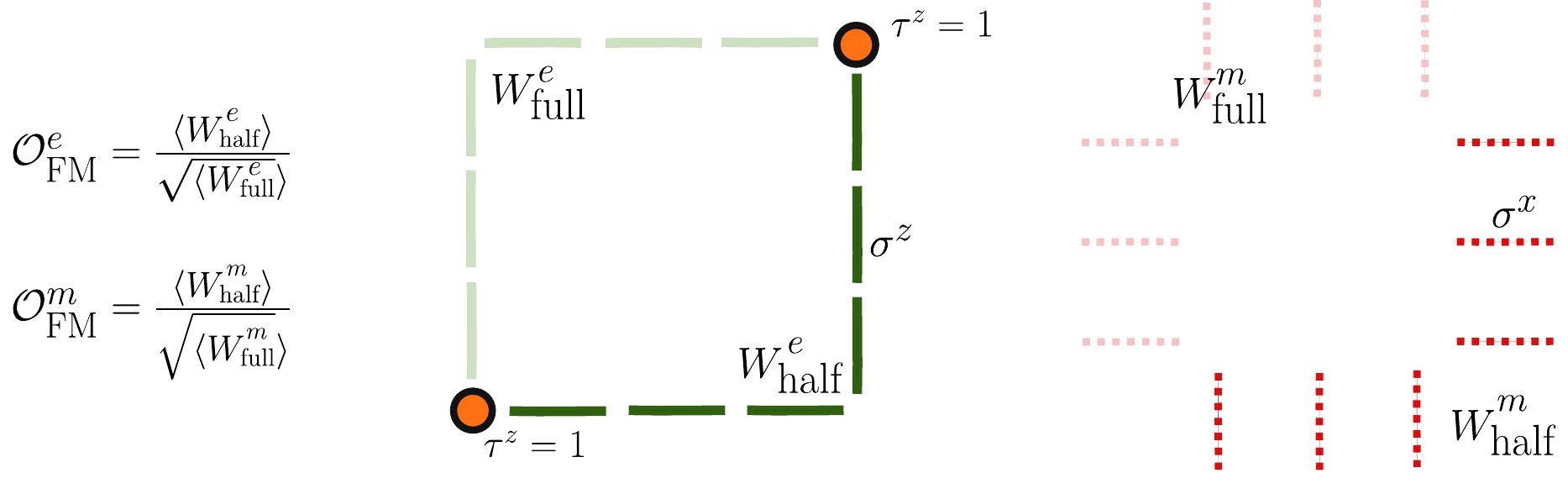}
	\caption{Depiction of the electric and magnetic Fredenhagen-Marcu order parameters. On the right, we show the Wilson ('t Hooft) lines and loops that enter the definitions.}
	\label{supfig:FMs_def}
\end{figure}

The precise lattice definitions of the electric and magnetic Fredenhagen-Marcu order parameters that we use in the main text, are provided graphically in \cref{supfig:FMs_def}. While both order parameters are in principle well defined away from the limiting cases $h_x=0$ ($h_z=0$), $\mathcal{O}_{\text{\text{FM}}}^e$ is designed to detect charge condensation and therefore it is more suitable as an order parameter for the Higgs phase. The opposite holds for $\mathcal{O}_{\text{\text{FM}}}^m$.

\begin{figure}[ht]
\captionsetup[subfigure]{labelformat=empty}
    \subfloat[\label{subfig:fms_1}]{}
    \subfloat[\label{subfig:fms_2}]{}
	\includegraphics[width=0.95\linewidth]{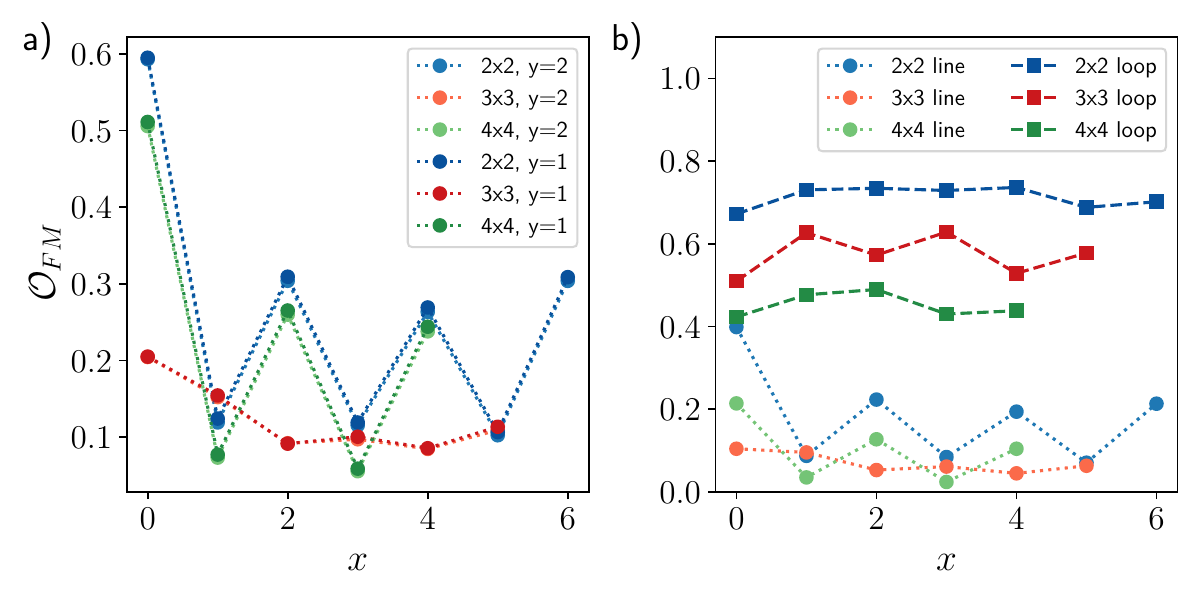}
	\caption{(a) Profile of the 4x4 FM order parameters along the horizontal direction on the self-dual line, at $h=0.8$ (VBS phase), for a cylinder with $L_y=8$. We plot the observable for two different values of $y$, showing the translational invariance along the periodic direction. The $x$-dependence shows, besides boundary effects, the presence of a size two unit cell. Odd-even effects (in the side-length of the loop) are evident. (b) We show separately the values of the numerator and denominator of expression \eqref{eq:FM_def}. The staggering is manifest either in the line or in the loop, depending on the odd-even size.}
	\label{supfig:FMs_all}
\end{figure}

In the main text, we show how the Fredenhagen-Marcu order parameter provides a clear indication of deconfinement transitions, both on the self-dual line and away from it. Since we study systems that are finite and have open boundary conditions in the horizontal direction, the FM order parameter, as any observable, is position-dependent. To mitigate finite-size effects, all the results displayed in the main text are obtained by computing the FM order parameter in the central portion of the system. For completeness, in \cref{subfig:fms_1}, we depict its values for all horizontal position $j$ (corresponding to the bottom left corner of the loop), showing the extent of finite size effect, the translational invariance along the periodic direction, and the significance of even-odd effects. In \cref{subfig:fms_2}, we show the line (numerator) and loop (denominator) contributions to the order parameter, which show how the spatial dependence is more prominent in either of them depending on even-odd effects.

\section{Response to a symmetry-breaking perturbation}
\label{app:duality_sus}
\begin{figure}[h]
\captionsetup[subfigure]{labelformat=empty}
\subfloat[\label{supfig:deltaJ_response_odd}]{}
\subfloat[\label{supfig:deltaJ_response_even}]{}
\includegraphics[width=\columnwidth]{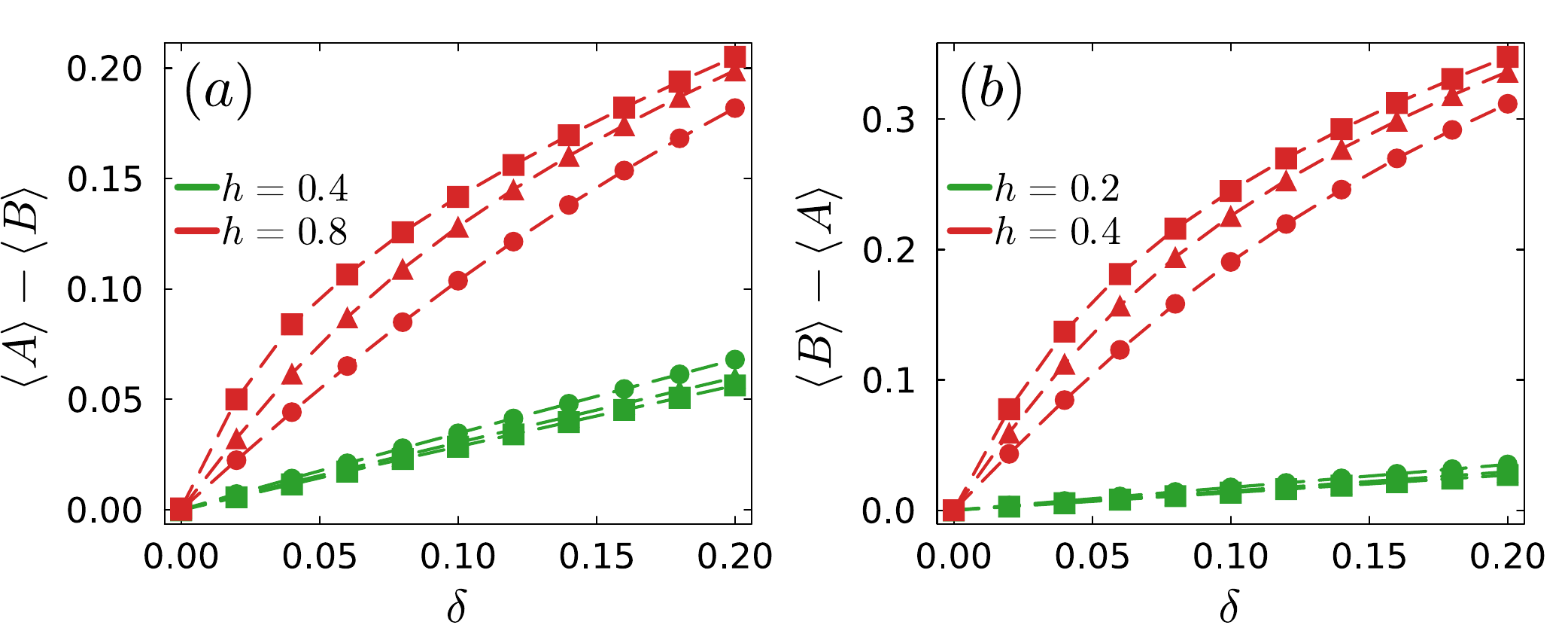}
	\caption{Difference between the average star and the plaquette terms as a function of the perturbative symmetry-breaking coupling near self-dual points with $h=h_{x}=h_{z}$ for (a) the odd Toric Code in the deconfined ($h=0.4$) and the VBS regions ($h=0.8$) and (b) the even Toric Code in the deconfined ($h=0.2$) and the symmetry broken segments ($h=0.4$) of the self-dual line. The circles, triangles and squares represent systems with $L=4,6$, and $8$, respectively.}
	\label{supfig:deltaJ_response}
\end{figure}

To further scrutinize our claim of self-duality symmetry-breaking order along the self-dual line for $h>h_{c}$, in \cref{supfig:deltaJ_response} we study the difference between the expectation values of the star and the plaquette terms as a function of symmetry-breaking perturbation, $\delta$, as defined in \cref{eq:deltaJ}. In the self-duality broken phase, the plaquette-vertex energy difference is expected to jump to a finite value for an infinitesimal $\delta$, resulting from a first-order transition across the self-dual line. On the contrary, one expects the plaquette-vertex energy difference to scale linearly with $\delta$ in the deconfined phase due to the absence of symmetry breaking. This is similar to the more standard symmetry-breaking phenomenon, such as in the Ising model, where introducing an infinitesimal symmetry-breaking field in the ordered phase results in a finite magnetization jump. On the contrary, a smooth evolution as a function of the external field is expected in a disordered phase.
\begin{figure}[t]
\captionsetup[subfigure]{labelformat=empty}
\vspace{0.2cm}
\includegraphics[width=0.9\linewidth]{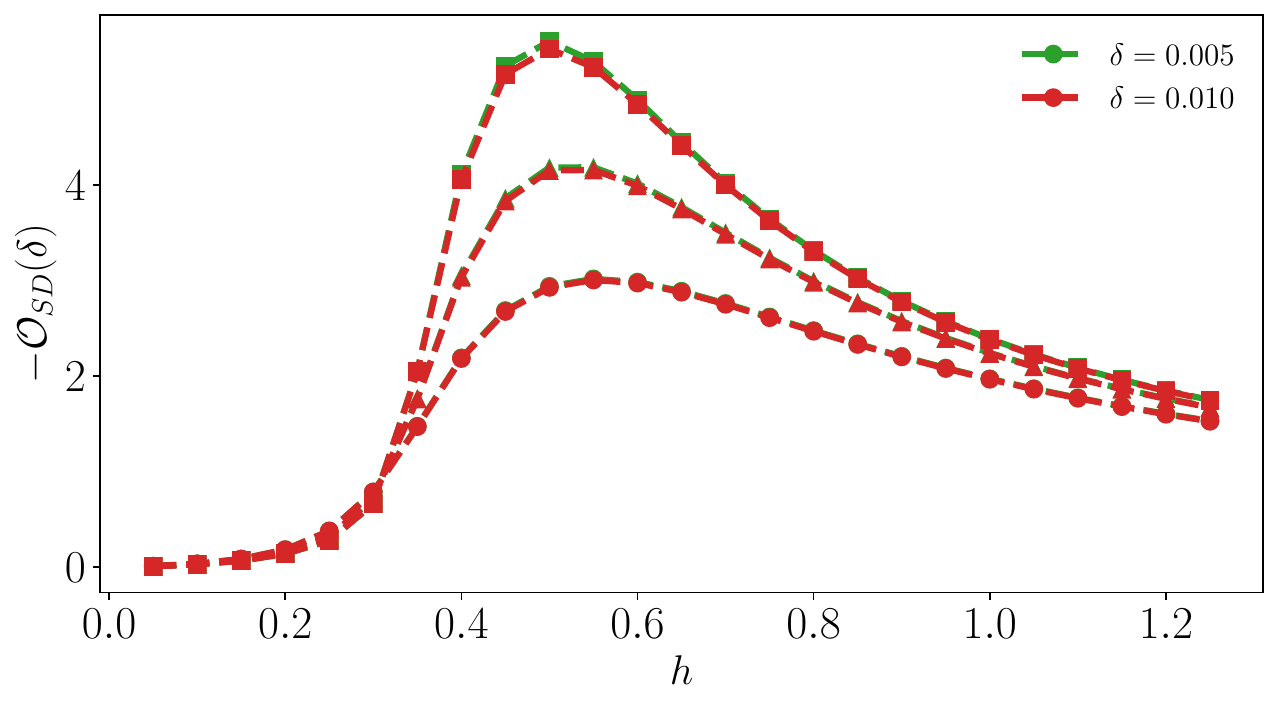}
	\caption{The self-duality symmetry-breaking susceptibility along the self-dual line for two different values of the symmetry-breaking parameter for the even Toric Code case. The circles, triangles and squares represent systems with $L=4,6,8$ respectively.}
	\label{supfig:sus_even}
\end{figure}

In \cref{supfig:deltaJ_response} we showcase the aforementioned behavior in the deconfined and self-duality broken phases for both the odd (\cref{supfig:deltaJ_response_odd}) and the even (\cref{supfig:deltaJ_response_even}) Toric Code models. Evidently, in the self-duality broken region, the plaquette-vertex energy difference sharpens around $\delta J =0$, suggesting a discontinuous jump in the thermodynamic limit. On the other hand, in the deconfined phase, the response scales linearly with $\delta $. 

Lastly, we measure $\mathcal{O}_{SD}(\delta )$ as shown for the odd Toric Code in \cref{subfig:SDB_SD} but for the even case in \cref{supfig:sus_even}. Here again, we observe that $\mathcal{O}_{SD}(\delta )$ transitions from a zero thermodynamic value in the deconfined phase with $h \lesssim 0.34$ to a behavior that diverges with system size in the self-duality broken region with $h \gtrsim 0.34$. This agreement with the well-established literature \cite{xu2024critical} for the even Toric Code supports our approach to resolving the phase diagram for the odd case.

\section{Numerical Details}
\label{app:numerics}

\begin{figure}[ht]
\captionsetup[subfigure]{labelformat=empty}
\includegraphics[width=\columnwidth]{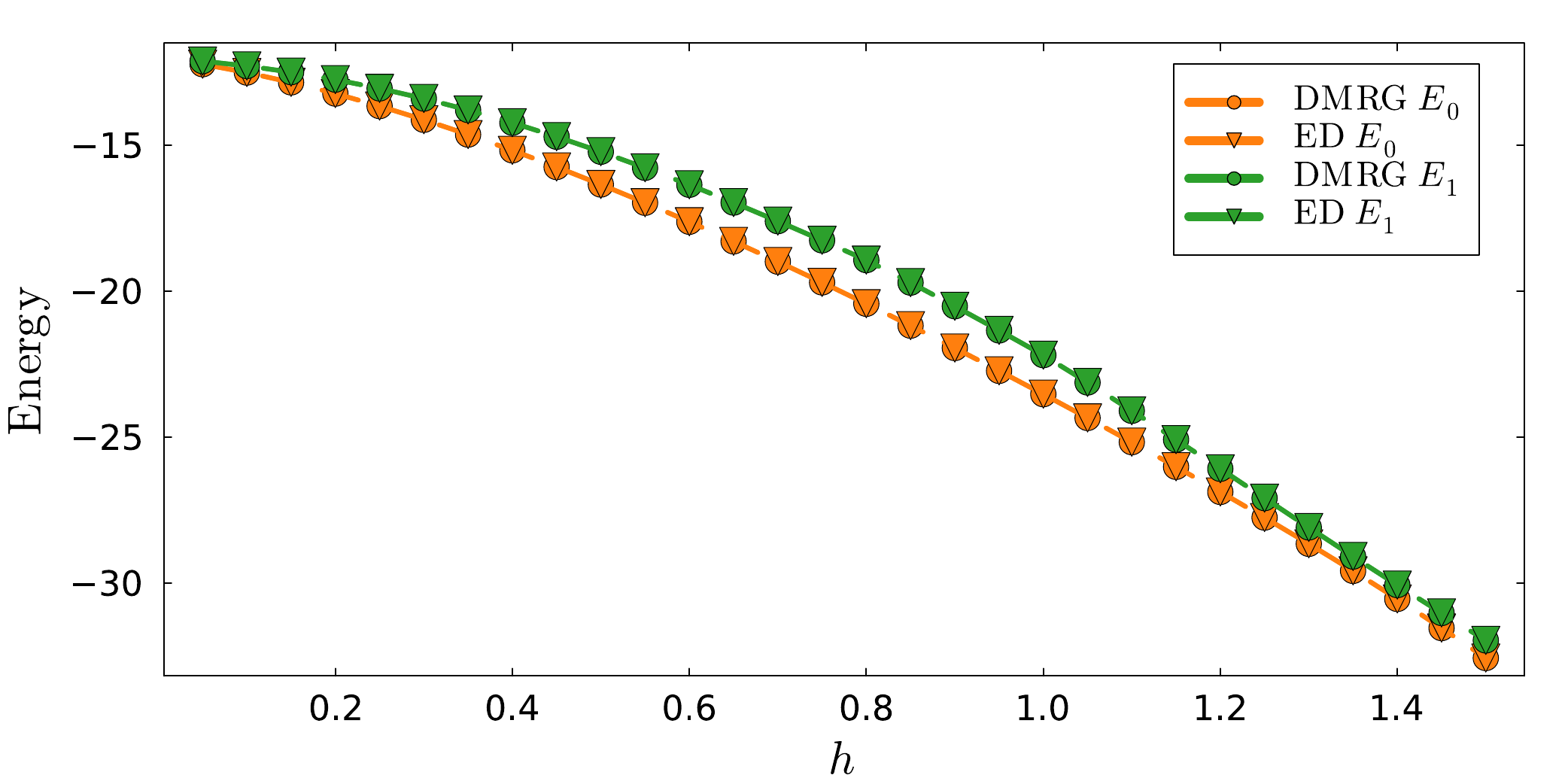}
	\caption{Comparison between ground state and first excited state energies obtained from Exact Diagonalization against our DMRG calculation along the self-dual line for an $L_{x}=3+1$, $L_y=2$ cylinder.}
	\label{supfig:Benchmark}
\end{figure}

To benchmark our Hamiltonian MPO implementation, we compare the DMRG ground state and first excited state energies to exact diagonalization calculations on a $L_{x}=3+1$, $L_y=2$ cylinder in \cref{supfig:Benchmark}. We observe perfect agreement between both methods.

\begin{figure}[h]
\captionsetup[subfigure]{labelformat=empty}
\includegraphics[width=\columnwidth]{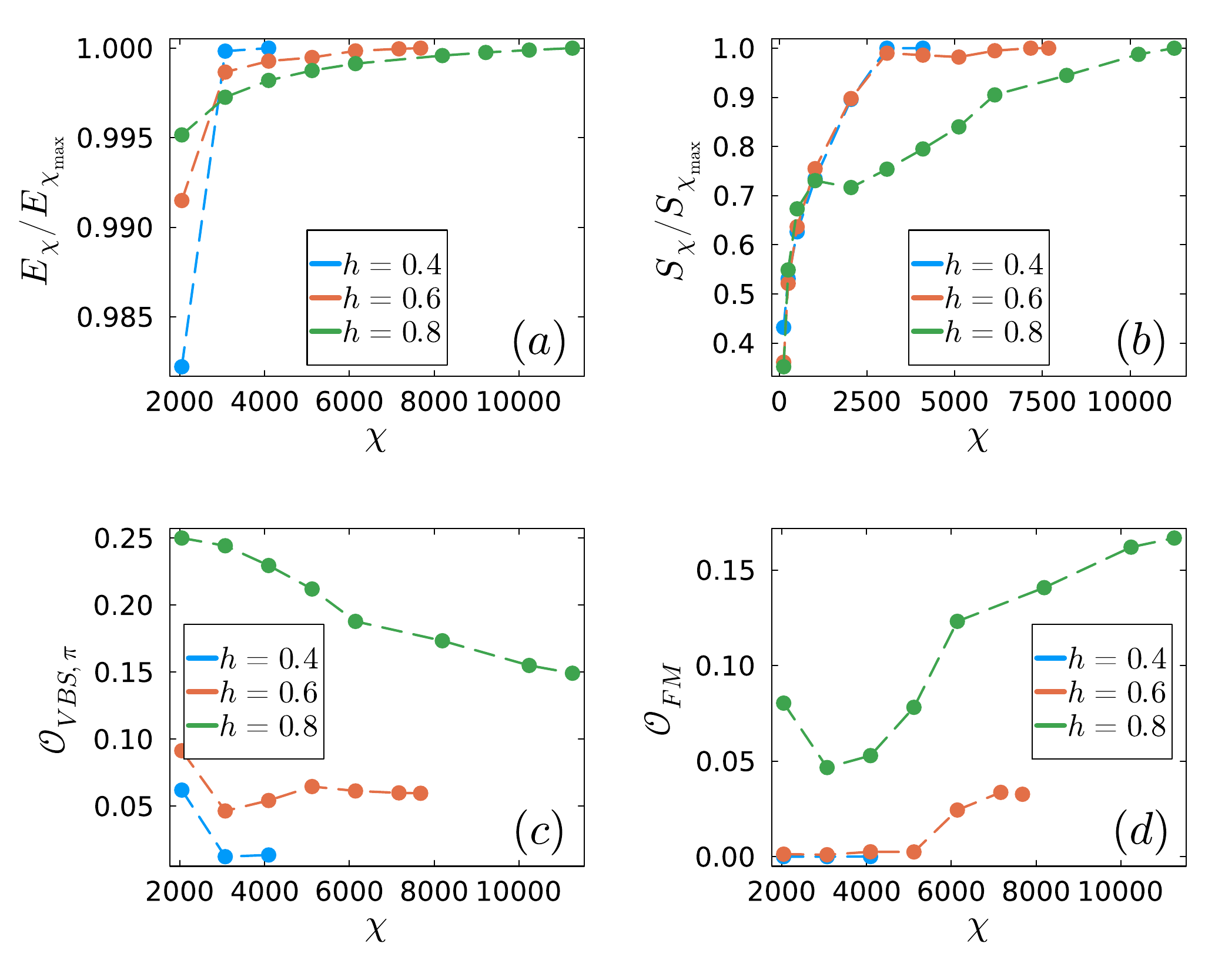}
	\caption{Convergence of (a) the energy and (b) the entropy (c) the VBS order parameter and (d) the FM order parameter at different points along the self-dual line as a function of the MPS bond dimension $\chi$ for an $L=10$ system.}
	\label{supfig:convergence}
\end{figure}

Additionally, to ensure proper convergence with the bond dimension, we track the flow of various observables as a function of bond dimension. In \cref{supfig:convergence}, we showcase the same for the hardest-to-compute $L=10$ system size up to the largest bond dimension ($\chi=11264$) achievable by our compute hardware. We see that while the observables are fully converged for $h=0.4,0.6$. The $h=0.8$ data, however, while not fully converged, shows signs of approaching its asymptotic value. We estimate that these numbers are close enough to the true value to accurately capture the qualitative behavior of the wavefunction.

\bibliography{library}

@misc{Zenodo,
      title  = {Zenodo repository with the numerical data for ``Odd Toric Code in a tilted field: Higgs-confinement multicriticality, spontaneous self-duality symmetry breaking, and valence bond solids},
      url = {https://zenodo.org/records/16686366}
    }

@article{Nahum_2024,
  title = {Worldsheet patching, 1-form symmetries, and ${\mathrm{Landau}}^{*}$ phase transitions},
  author = {Serna, Pablo and Somoza, Andr\'es M. and Nahum, Adam},
  journal = {Phys. Rev. B},
  volume = {110},
  issue = {11},
  pages = {115102},
  numpages = {23},
  year = {2024},
  month = {Sep},
  publisher = {American Physical Society},
  doi = {10.1103/PhysRevB.110.115102},
  url = {https://link.aps.org/doi/10.1103/PhysRevB.110.115102}
}

@article{Scaffidi_2021,
  title = {Evidence for deconfined $U(1)$ gauge theory at the transition between toric code and double semion},
  author = {Dupont, Maxime and Gazit, Snir and Scaffidi, Thomas},
  journal = {Phys. Rev. B},
  volume = {103},
  issue = {14},
  pages = {L140412},
  numpages = {7},
  year = {2021},
  month = {Apr},
  publisher = {American Physical Society},
  doi = {10.1103/PhysRevB.103.L140412},
  url = {https://link.aps.org/doi/10.1103/PhysRevB.103.L140412}
}

@misc{dumitrescu_2026,
      title={From QED$_3$ to Self-Dual Multicriticality in the Fradkin-Shenker Model}, 
      author={Thomas T. Dumitrescu and Pierluigi Niro and Ryan Thorngren},
      year={2026},
      eprint={2602.23420},
      archivePrefix={arXiv},
      primaryClass={cond-mat.str-el},
      url={https://arxiv.org/abs/2602.23420}, 
}

@misc{jian_2017,
      title={Emergent Symmetry and Tricritical Points near the deconfined Quantum Critical Point}, 
      author={Chao-Ming Jian and Alex Rasmussen and Yi-Zhuang You and Cenke Xu},
      year={2017},
      eprint={1708.03050},
      archivePrefix={arXiv},
      primaryClass={cond-mat.str-el},
      url={https://arxiv.org/abs/1708.03050}, 
}

@article{Elitzur_1975,
  title = {Impossibility of spontaneously breaking local symmetries},
  author = {Elitzur, S.},
  journal = {Phys. Rev. D},
  volume = {12},
  issue = {12},
  pages = {3978--3982},
  numpages = {0},
  year = {1975},
  month = {Dec},
  publisher = {American Physical Society},
  doi = {10.1103/PhysRevD.12.3978},
  url = {https://link.aps.org/doi/10.1103/PhysRevD.12.3978}
}

@book{sachdev_2023, 
        place={Cambridge}, 
        title={Quantum Phases of Matter}, 
        DOI={10.1017/9781009212717}, 
        publisher={Cambridge University Press}, 
        author={Sachdev, Subir}, 
        year={2023}
}

@article{KITAEV2003,
title = {Fault-tolerant quantum computation by anyons},
journal = {Annals of Physics},
volume = {303},
number = {1},
pages = {2-30},
year = {2003},
issn = {0003-4916},
doi = {https://doi.org/10.1016/S0003-4916(02)00018-0},
url = {https://www.sciencedirect.com/science/article/pii/S0003491602000180},
author = {A.Yu. Kitaev}
}

@article{Borla_2024,
  title = {Deconfined quantum criticality in Ising gauge theory entangled with single-component fermions},
  author = {Borla, Umberto and Gazit, Snir and Moroz, Sergej},
  journal = {Phys. Rev. B},
  volume = {110},
  issue = {20},
  pages = {L201110},
  numpages = {6},
  year = {2024},
  month = {Nov},
  publisher = {American Physical Society},
  doi = {10.1103/PhysRevB.110.L201110},
  url = {https://link.aps.org/doi/10.1103/PhysRevB.110.L201110}
}

@article{Gazit_2020,
  title = {Fermi Surface Reconstruction without Symmetry Breaking},
  author = {Gazit, Snir and Assaad, Fakher F. and Sachdev, Subir},
  journal = {Phys. Rev. X},
  volume = {10},
  issue = {4},
  pages = {041057},
  numpages = {15},
  year = {2020},
  month = {Dec},
  publisher = {American Physical Society},
  doi = {10.1103/PhysRevX.10.041057},
  url = {https://link.aps.org/doi/10.1103/PhysRevX.10.041057}
}

@article{Elio_2024,
  title = {$\mathbb{Z}_{N}$ lattice gauge theories with matter fields},
  author = {Roy, Kaustubh and K\"onig, Elio J.},
  journal = {Phys. Rev. B},
  volume = {109},
  issue = {19},
  pages = {195108},
  numpages = {22},
  year = {2024},
  month = {May},
  publisher = {American Physical Society},
  doi = {10.1103/PhysRevB.109.195108},
  url = {https://link.aps.org/doi/10.1103/PhysRevB.109.195108}
}

@article{Meng_2021,
  title = {Fermi arcs and pseudogap in a lattice model of a doped orthogonal metal},
  author = {Chen, Chuang and Yuan, Tian and Qi, Yang and Meng, Zi Yang},
  journal = {Phys. Rev. B},
  volume = {103},
  issue = {16},
  pages = {165131},
  numpages = {12},
  year = {2021},
  month = {Apr},
  publisher = {American Physical Society},
  doi = {10.1103/PhysRevB.103.165131},
  url = {https://link.aps.org/doi/10.1103/PhysRevB.103.165131}
}

@article{Tsvelik_2020,
  title = {Soluble limit and criticality of fermions in ${\mathbb{Z}}_{2}$ gauge theories},
  author = {K\"onig, Elio J. and Coleman, Piers and Tsvelik, Alexei M.},
  journal = {Phys. Rev. B},
  volume = {102},
  issue = {15},
  pages = {155143},
  numpages = {15},
  year = {2020},
  month = {Oct},
  publisher = {American Physical Society},
  doi = {10.1103/PhysRevB.102.155143},
  url = {https://link.aps.org/doi/10.1103/PhysRevB.102.155143}
}

@ARTICLE{Goldstein_2024,
       author = {{Feldman}, Noa and {Knaute}, Johannes and {Zohar}, Erez and {Goldstein}, Moshe},
        title = "{Superselection-resolved entanglement in lattice gauge theories: a tensor network approach}",
      journal = {Journal of High Energy Physics},
         year = 2024,
        month = may,
       volume = {2024},
       number = {5},
          eid = {83},
        pages = {83},
          doi = {10.1007/JHEP05(2024)083},
 primaryClass = {quant-ph}
}

@article{Wigner_1970,
  title = {Superselection Rule for Charge},
  author = {Wick, G. -C. and Wightman, A. S. and Wigner, Eugene P.},
  journal = {Phys. Rev. D},
  volume = {1},
  issue = {12},
  pages = {3267--3269},
  numpages = {0},
  year = {1970},
  month = {Jun},
  publisher = {American Physical Society},
  doi = {10.1103/PhysRevD.1.3267},
  url = {https://link.aps.org/doi/10.1103/PhysRevD.1.3267}
}

@misc{linsel_2025,
      title={Independent e- and m-anyon confinement in the parallel field toric code on non-square lattices}, 
      author={Simon M. Linsel and Lode Pollet and Fabian Grusdt},
      year={2025},
      eprint={2504.03512},
      archivePrefix={arXiv},
      primaryClass={quant-ph},
      url={https://arxiv.org/abs/2504.03512}, 
}

@article{Zohar_2022,
author = {Zohar, Erez },
title = {Quantum simulation of lattice gauge theories in more than one space dimension—requirements, challenges and methods},
journal = {Philosophical Transactions of the Royal Society A: Mathematical, Physical and Engineering Sciences},
volume = {380},
number = {2216},
pages = {20210069},
year = {2022},
doi = {10.1098/rsta.2021.0069},

URL = {https://royalsocietypublishing.org/doi/abs/10.1098/rsta.2021.0069}
}

@article{Jalabert_1991,
  title = {Spontaneous alignment of frustrated bonds in an anisotropic, three-dimensional Ising model},
  author = {Jalabert, Rodolfo A. and Sachdev, Subir},
  journal = {Phys. Rev. B},
  volume = {44},
  issue = {2},
  pages = {686--690},
  numpages = {0},
  year = {1991},
  month = {Jul},
  publisher = {American Physical Society},
  doi = {10.1103/PhysRevB.44.686},
  url = {https://link.aps.org/doi/10.1103/PhysRevB.44.686}
}

@book{wilczek1990fractional,
  title={Fractional statistics and anyon superconductivity},
  author={Wilczek, Frank},
  volume={5},
  year={1990},
  publisher={World scientific}
}

@article{Nahum2020,
  title = {Self-Dual Criticality in Three-Dimensional $\mathbb{Z}_2$ Gauge Theory with Matter},
  author = {Somoza, Andr\'es M. and Serna, Pablo and Nahum, Adam},
  journal = {Phys. Rev. X},
  volume = {11},
  issue = {4},
  pages = {041008},
  numpages = {32},
  year = {2021},
  month = {Oct},
  publisher = {American Physical Society},
  doi = {10.1103/PhysRevX.11.041008},
  url = {https://link.aps.org/doi/10.1103/PhysRevX.11.041008}
}

@article{vidal2009,
  title = {Low-energy effective theory of the toric code model in a parallel magnetic field},
  author = {Vidal, Julien and Dusuel, S\'ebastien and Schmidt, Kai Phillip},
  journal = {Phys. Rev. B},
  volume = {79},
  issue = {3},
  pages = {033109},
  numpages = {4},
  year = {2009},
  month = {Jan},
  publisher = {American Physical Society},
  doi = {10.1103/PhysRevB.79.033109},
  url = {https://link.aps.org/doi/10.1103/PhysRevB.79.033109}
}

@article{borla2022,
  title = {Quantum phases of two-dimensional $\mathbb{Z}_{2}$ gauge theory coupled to single-component fermion matter},
  author = {Borla, Umberto and Jeevanesan, Bhilahari and Pollmann, Frank and Moroz, Sergej},
  journal = {Phys. Rev. B},
  volume = {105},
  issue = {7},
  pages = {075132},
  numpages = {20},
  year = {2022},
  month = {Feb},
  publisher = {American Physical Society},
  doi = {10.1103/PhysRevB.105.075132},
  url = {https://link.aps.org/doi/10.1103/PhysRevB.105.075132}
}

@misc{kufel2024,
      title={Approximately-symmetric neural networks for quantum spin liquids}, 
      author={Dominik S. Kufel and Jack Kemp and Simon M. Linsel and Chris R. Laumann and Norman Y. Yao},
      year={2024},
      eprint={2405.17541},
      archivePrefix={arXiv},
      primaryClass={quant-ph},
      url={https://arxiv.org/abs/2405.17541}, 
}

@article{Kogut_1979,
  title = {An introduction to lattice gauge theory and spin systems},
  author = {Kogut, John B.},
  journal = {Rev. Mod. Phys.},
  volume = {51},
  issue = {4},
  pages = {659--713},
  numpages = {0},
  year = {1979},
  month = {Oct},
  publisher = {American Physical Society},
  doi = {10.1103/RevModPhys.51.659},
  url = {https://link.aps.org/doi/10.1103/RevModPhys.51.659}
}

@Article{tenpy2024,
    title={{Tensor network Python (TeNPy) version 1}},
    author={Johannes Hauschild and Jakob Unfried and Sajant Anand and Bartholomew Andrews and Marcus Bintz and Umberto Borla and Stefan Divic and Markus Drescher and Jan Geiger and Martin Hefel and Kévin Hémery and Wilhelm Kadow and Jack Kemp and Nico Kirchner and Vincent S. Liu and Gunnar Möller and Daniel Parker and Michael Rader and Anton Romen and Samuel Scalet and Leon Schoonderwoerd and Maximilian Schulz and Tomohiro Soejima and Philipp Thoma and Yantao Wu and Philip Zechmann and Ludwig Zweng and Roger S. K. Mong and Michael P. Zaletel and Frank Pollmann},
    journal={SciPost Phys. Codebases},
    pages={41},
    year={2024},
    publisher={SciPost},
    doi={10.21468/SciPostPhysCodeb.41},
    url={https://scipost.org/10.21468/SciPostPhysCodeb.41},
}

@article{ITensor,
	title={{The ITensor Software Library for Tensor Network Calculations}},
	author={Matthew Fishman and Steven R. White and E. Miles Stoudenmire},
	journal={SciPost Phys. Codebases},
	pages={4},
	year={2022},
	publisher={SciPost},
	doi={10.21468/SciPostPhysCodeb.4},
	url={https://scipost.org/10.21468/SciPostPhysCodeb.4},
}

@Article{QuSpin,
	title={{QuSpin: a Python package for dynamics and exact diagonalisation of quantum many body systems. Part II: bosons, fermions and higher spins}},
	author={Phillip Weinberg and Marin Bukov},
	journal={SciPost Phys.},
	volume={7},
	pages={020},
	year={2019},
	publisher={SciPost},
	doi={10.21468/SciPostPhys.7.2.020},
	url={https://scipost.org/10.21468/SciPostPhys.7.2.020},
}

@article{ringel2024,
  title = {Machine learning the operator content of the critical self-dual Ising-Higgs lattice gauge theory},
  author = {Oppenheim, Lior and Koch-Janusz, Maciej and Gazit, Snir and Ringel, Zohar},
  journal = {Phys. Rev. Res.},
  volume = {6},
  issue = {4},
  pages = {043322},
  numpages = {10},
  year = {2024},
  month = {Dec},
  publisher = {American Physical Society},
  doi = {10.1103/PhysRevResearch.6.043322},
  url = {https://link.aps.org/doi/10.1103/PhysRevResearch.6.043322}
}

@misc{Laumann_2025,
      title={Hall-on-Toric: Descendant Laughlin state in the chiral $\mathbb{Z}_p$ toric code}, 
      author={Robin Schäfer and Claudio Chamon and Chris R. Laumann},
      year={2025},
      eprint={2507.02035},
      archivePrefix={arXiv},
      primaryClass={cond-mat.str-el},
      url={https://arxiv.org/abs/2507.02035}, 
}

@article{Carleo_2017,
   title={Solving the quantum many-body problem with artificial neural networks},
   volume={355},
   ISSN={1095-9203},
   url={http://dx.doi.org/10.1126/science.aag2302},
   DOI={10.1126/science.aag2302},
   number={6325},
   journal={Science},
   publisher={American Association for the Advancement of Science (AAAS)},
   author={Carleo, Giuseppe and Troyer, Matthias},
   year={2017},
   month=feb, pages={602–606} }

@misc{Bohrdt_2024,
      title={From Architectures to Applications: A Review of Neural Quantum States}, 
      author={Hannah Lange and Anka Van de Walle and Atiye Abedinnia and Annabelle Bohrdt},
      year={2024},
      eprint={2402.09402},
      archivePrefix={arXiv},
      primaryClass={cond-mat.dis-nn},
      url={https://arxiv.org/abs/2402.09402}, 
}

@article{mila2012,
  title = {Evidence for Columnar Order in the Fully Frustrated Transverse Field Ising Model on the Square Lattice},
  author = {Wenzel, Sandro and Coletta, Tommaso and Korshunov, Sergey E. and Mila, Fr\'ed\'eric},
  journal = {Phys. Rev. Lett.},
  volume = {109},
  issue = {18},
  pages = {187202},
  numpages = {5},
  year = {2012},
  month = {Nov},
  publisher = {American Physical Society},
  doi = {10.1103/PhysRevLett.109.187202},
  url = {https://link.aps.org/doi/10.1103/PhysRevLett.109.187202}
}

@article{barbiero2019,
	uthor = {Luca Barbiero  and Christian Schweizer  and Monika Aidelsburger  and Eugene Demler  and Nathan Goldman  and Fabian Grusdt },
title = {Coupling ultracold matter to dynamical gauge fields in optical lattices: From flux attachment to $\mathbb{Z}_{2}$ lattice gauge theories},
journal = {Science Advances},
volume = {5},
number = {10},
pages = {eaav7444},
year = {2019},
doi = {10.1126/sciadv.aav7444},
URL = {https://www.science.org/doi/abs/10.1126/sciadv.aav7444}}

@article{Homeier_2021,
  title = {$\mathbb{Z}_{2}$ lattice gauge theories and Kitaev's toric code: A scheme for analog quantum simulation},
  author = {Homeier, Lukas and Schweizer, Christian and Aidelsburger, Monika and Fedorov, Arkady and Grusdt, Fabian},
  journal = {Phys. Rev. B},
  volume = {104},
  issue = {8},
  pages = {085138},
  numpages = {19},
  year = {2021},
  month = {Aug},
  publisher = {American Physical Society},
  doi = {10.1103/PhysRevB.104.085138},
  url = {https://link.aps.org/doi/10.1103/PhysRevB.104.085138}
}

@article{mcgreevy2023generalized,
  title={Generalized symmetries in condensed matter},
  author={McGreevy, John},
  journal={Annual Review of Condensed Matter Physics},
  volume={14},
  pages={57--82},
  year={2023},
  publisher={Annual Reviews}
}

@article{gaiotto2015generalized,
   title={Generalized global symmetries},
   volume={2015},
   ISSN={1029-8479},
   url={http://dx.doi.org/10.1007/JHEP02(2015)172},
   DOI={10.1007/jhep02(2015)172},
   number={2},
   journal={Journal of High Energy Physics},
   publisher={Springer Science and Business Media LLC},
   author={Gaiotto, Davide and Kapustin, Anton and Seiberg, Nathan and Willett, Brian},
   year={2015},
   month=feb }

@article{Lumia_2022,
  title = {Two-Dimensional ${\mathbb{Z}}_{2}$ Lattice Gauge Theory on a Near-Term Quantum Simulator: Variational Quantum Optimization, Confinement, and Topological Order},
  author = {Lumia, Luca and Torta, Pietro and Mbeng, Glen B. and Santoro, Giuseppe E. and Ercolessi, Elisa and Burrello, Michele and Wauters, Matteo M.},
  journal = {PRX Quantum},
  volume = {3},
  issue = {2},
  pages = {020320},
  numpages = {22},
  year = {2022},
  month = {Apr},
  publisher = {American Physical Society},
  doi = {10.1103/PRXQuantum.3.020320},
  url = {https://link.aps.org/doi/10.1103/PRXQuantum.3.020320}
}

@article{Reinis2023,
  title = {Quantum simulation of $\mathbb{Z}_{2}$ lattice gauge theory with minimal resources},
  author = {Irmejs, Reinis and Ba\~nuls, Mari-Carmen and Cirac, J. Ignacio},
  journal = {Phys. Rev. D},
  volume = {108},
  issue = {7},
  pages = {074503},
  numpages = {11},
  year = {2023},
  month = {Oct},
  publisher = {American Physical Society},
  doi = {10.1103/PhysRevD.108.074503},
  url = {https://link.aps.org/doi/10.1103/PhysRevD.108.074503}
}

@Article{Homeier2023,
author={Homeier, Lukas
and Bohrdt, Annabelle
and Linsel, Simon
and Demler, Eugene
and Halimeh, Jad C.
and Grusdt, Fabian},
title = {Realistic Scheme for Quantum Simulation of {$\mathbb{Z}_2$} Lattice Gauge Theories with Dynamical Matter in (2+1)D},
journal={Communications Physics},
year={2023},
month={Jun},
day={05},
volume={6},
number={1},
pages={127},
issn={2399-3650},
doi={10.1038/s42005-023-01237-6},
url={https://doi.org/10.1038/s42005-023-01237-6}
}

@article{Senthil_2000,
  title = {${Z}_{2}$ gauge theory of electron fractionalization in strongly correlated systems},
  author = {Senthil, T. and Fisher, Matthew P. A.},
  journal = {Phys. Rev. B},
  volume = {62},
  issue = {12},
  pages = {7850--7881},
  numpages = {0},
  year = {2000},
  month = {Sep},
  publisher = {American Physical Society},
  doi = {10.1103/PhysRevB.62.7850},
  url = {https://link.aps.org/doi/10.1103/PhysRevB.62.7850}
}

@article{gazit2018,
author = {Snir Gazit  and Fakher F. Assaad  and Subir Sachdev  and Ashvin Vishwanath  and Chong Wang },
title = {Confinement transition of $\mathbb{Z}_2$ gauge theories coupled to massless fermions: Emergent quantum chromodynamics and ${SO}(5)$ symmetry},
journal = {Proceedings of the National Academy of Sciences},
volume = {115},
number = {30},
pages = {E6987-E6995},
year = {2018},
doi = {10.1073/pnas.1806338115},
URL = {https://www.pnas.org/doi/abs/10.1073/pnas.1806338115}
}

@article{bonati2022,
  title = {Multicritical point of the three-dimensional $\mathbb{Z}_2$ gauge {H}iggs model},
  author = {Bonati, Claudio and Pelissetto, Andrea and Vicari, Ettore},
  journal = {Phys. Rev. B},
  volume = {105},
  issue = {16},
  pages = {165138},
  numpages = {9},
  year = {2022},
  month = {Apr},
  publisher = {American Physical Society},
  doi = {10.1103/PhysRevB.105.165138},
  url = {https://link.aps.org/doi/10.1103/PhysRevB.105.165138}
}

@article{tupitsyn2010,
  title = {Topological multicritical point in the phase diagram of the toric code model and three-dimensional lattice gauge {H}iggs model},
  author = {Tupitsyn, I. S. and Kitaev, A. and Prokof'ev, N. V. and Stamp, P. C. E.},
  journal = {Phys. Rev. B},
  volume = {82},
  issue = {8},
  pages = {085114},
  numpages = {5},
  year = {2010},
  month = {Aug},
  publisher = {American Physical Society},
  doi = {10.1103/PhysRevB.82.085114},
  url = {https://link.aps.org/doi/10.1103/PhysRevB.82.085114}
}

@article{Verresen_2021,
  title = {Prediction of Toric Code Topological Order from Rydberg Blockade},
  author = {Verresen, Ruben and Lukin, Mikhail D. and Vishwanath, Ashvin},
  journal = {Phys. Rev. X},
  volume = {11},
  issue = {3},
  pages = {031005},
  numpages = {23},
  year = {2021},
  month = {Jul},
  publisher = {American Physical Society},
  doi = {10.1103/PhysRevX.11.031005},
  url = {https://link.aps.org/doi/10.1103/PhysRevX.11.031005}
}

@article{Assaad_2016,
  title = {Simple Fermionic Model of Deconfined Phases and Phase Transitions},
  author = {Assaad, F. F. and Grover, Tarun},
  journal = {Phys. Rev. X},
  volume = {6},
  issue = {4},
  pages = {041049},
  numpages = {19},
  year = {2016},
  month = {Dec},
  publisher = {American Physical Society},
  doi = {10.1103/PhysRevX.6.041049},
  url = {https://link.aps.org/doi/10.1103/PhysRevX.6.041049}
}

@article{gazit2017,
	doi = {10.1038/nphys4028},
	url = {https://doi.org/10.1038%2Fnphys4028},
	year = 2017,
	month = {feb},
	publisher = {Springer Science and Business Media {LLC}
},
	volume = {13},
	number = {5},
	pages = {484--490},
	author = {Snir Gazit and Mohit Randeria and Ashvin Vishwanath},
	title = {Emergent Dirac fermions and broken symmetries in confined and deconfined phases of $\mathbb{Z}_2$ gauge~theories},
	journal = {Nature Physics}
}

@misc{linsel2025independentemanyonconfinement,
      title={Independent e- and m-anyon confinement in the parallel field toric code on non-square lattices}, 
      author={Simon M. Linsel and Lode Pollet and Fabian Grusdt},
      year={2025},
      eprint={2504.03512},
      archivePrefix={arXiv},
      primaryClass={quant-ph},
      url={https://arxiv.org/abs/2504.03512}, 
}

@article{Moessner_2001,
  title = {Short-ranged resonating valence bond physics, quantum dimer models, and Ising gauge theories},
  author = {Moessner, R. and Sondhi, S. L. and Fradkin, Eduardo},
  journal = {Phys. Rev. B},
  volume = {65},
  issue = {2},
  pages = {024504},
  numpages = {16},
  year = {2001},
  month = {Dec},
  publisher = {American Physical Society},
  doi = {10.1103/PhysRevB.65.024504},
  url = {https://link.aps.org/doi/10.1103/PhysRevB.65.024504}
}

@article{Geraedts_2013,
   title={Exact realization of integer and fractional quantum Hall phases in models in},
   volume={334},
   ISSN={0003-4916},
   url={http://dx.doi.org/10.1016/j.aop.2013.03.017},
   DOI={10.1016/j.aop.2013.03.017},
   journal={Annals of Physics},
   publisher={Elsevier BV},
   author={Geraedts, Scott D. and Motrunich, Olexei I.},
   year={2013},
   month=jul, pages={288–315} }

@article{Gregor_2011,
	doi = {10.1088/1367-2630/13/2/025009},
	url = {https://doi.org/10.1088/1367-2630/13/2/025009},
	year = 2011,
	month = {feb},
	publisher = {{IOP} Publishing},
	volume = {13},
	number = {2},
	pages = {025009},
	author = {K Gregor and David A Huse and R Moessner and S L Sondhi},
	title = {Diagnosing deconfinement and topological order},
	journal = {New Journal of Physics}
}

@article{Wegner_1971,
    author = {Wegner, Franz J.},
    title = {Duality in Generalized Ising Models and Phase Transitions without Local Order Parameters},
    journal = {Journal of Mathematical Physics},
    volume = {12},
    number = {10},
    pages = {2259-2272},
    year = {1971},
    month = {10},
    issn = {0022-2488},
    doi = {10.1063/1.1665530},
    url = {https://doi.org/10.1063/1.1665530},
    eprint = {https://pubs.aip.org/aip/jmp/article-pdf/12/10/2259/19106483/2259\_1\_online.pdf},
}

@misc{verresen2022higgs,
      title={Higgs Condensates are Symmetry-Protected Topological Phases: I. Discrete Symmetries}, 
      author={Ruben Verresen and Umberto Borla and Ashvin Vishwanath and Sergej Moroz and Ryan Thorngren},
      year={2024},
      eprint={2211.01376},
      archivePrefix={arXiv},
      primaryClass={cond-mat.str-el},
      url={https://arxiv.org/abs/2211.01376}, 
}

@article{Fredenhagen_1986,
  title = {Confinement criterion for QCD with dynamical quarks},
  author = {Fredenhagen, Klaus and Marcu, Mihail},
  journal = {Phys. Rev. Lett.},
  volume = {56},
  issue = {3},
  pages = {223--224},
  numpages = {0},
  year = {1986},
  month = {Jan},
  publisher = {American Physical Society},
  doi = {10.1103/PhysRevLett.56.223},
  url = {https://link.aps.org/doi/10.1103/PhysRevLett.56.223}
}

@article{savary2016quantum,
   title={Quantum spin liquids: a review},
   volume={80},
   ISSN={1361-6633},
   url={http://dx.doi.org/10.1088/0034-4885/80/1/016502},
   DOI={10.1088/0034-4885/80/1/016502},
   number={1},
   journal={Reports on Progress in Physics},
   publisher={IOP Publishing},
   author={Savary, Lucile and Balents, Leon},
   year={2016},
   month=nov, pages={016502} }

@article{Fradkin_1979,
  title = {Phase diagrams of lattice gauge theories with {H}iggs fields},
  author = {Fradkin, Eduardo and Shenker, Stephen H.},
  journal = {Phys. Rev. D},
  volume = {19},
  issue = {12},
  pages = {3682--3697},
  numpages = {0},
  year = {1979},
  month = {Jun},
  publisher = {American Physical Society},
  doi = {10.1103/PhysRevD.19.3682},
  url = {https://link.aps.org/doi/10.1103/PhysRevD.19.3682}
}

@book{Fradkin2013,
	title        = {{Field Theories of Condensed Matter Physics}},
	author       = {Fradkin, E},
	year         = 2013,
	publisher    = {Cambridge University Press},
	isbn         = 9780521764445,
	url          = {http://books.google.com/books?hl=en\&lr=\&id=x7\_6MX4ye\_wC\&oi=fnd\&pg=PR11\&dq=Field+Theories+of+Condensed+Matter+Physics\&ots=OTMzLv0\_tI\&sig=iT7am\_ANJcpca8PXm7hBPm7Dl1E}
}

@book{wenbook,
	title        = {Quantum Field Theory of Many-Body Systems},
	author       = {Wen, X.G.},
	year         = 2004,
	publisher    = {OUP Oxford},
	series       = {Oxford Graduate Texts},
	isbn         = 9780198530947,
	url          = {https://books.google.com/books?id=llnlrfdR4YgC},
	lccn         = 2004301677
}

@misc{xu2024critical,
      title={Critical behavior of the {F}redenhagen-{M}arcu order parameter at topological phase transitions}, 
      author={Wen-Tao Xu and Frank Pollmann and Michael Knap},
      year={2024},
      eprint={2402.00127},
      archivePrefix={arXiv},
      primaryClass={cond-mat.str-el}
}
\end{document}